\documentclass[twocolumn]{aastex63x}
\usepackage[utf8]{inputenc}
\usepackage{amsmath}
\usepackage{xcolor}

\def\hubble{Hubble}
\def\hnot{$H_0$}
\def\kmsmpc{${\rm km\;s^{-1}\; Mpc^{-1}}$}
\def\hst{\textit{HST}}
\def\wfc{\textit{WFC3}}
\def\snia{SNe Ia}

\begin{document}

\title{Connecting Infrared Surface Brightness Fluctuation Distances to Type Ia Supernova Hosts: \\ Testing the Top Rung of the Distance Ladder}

\author[0000-0003-4069-2817]{Peter Garnavich}
\affiliation{University of Notre Dame, Department of Physics and Astronomy, 225 Nieuwland Science Hall, Notre Dame, IN 46556}

\author[0000-0003-4773-4602]{Charlotte M. Wood}
\affiliation{University of Notre Dame, Department of Physics and Astronomy, 225 Nieuwland Science Hall, Notre Dame, IN 46556}

\author[0000-0002-0370-157X]{Peter Milne}
\affiliation{University of Arizona, Steward Observatory, 933 N. Cherry Avenue, Tucson, AZ 85721, USA}

\author[0000-0001-8762-8906]{Joseph B. Jensen}
\affiliation{Utah Valley University, Department of Physics, 800 W. University Parkway, MS 179, Orem, UT 84058, USA}

\author[0000-0002-5213-3548]{John P. Blakeslee}
\affiliation{NSF's NOIRLab, 950 N. Cherry Ave., Tucson, AZ 85719, USA}

\author[0000-0001-6272-5507]{Peter J. Brown}
\affiliation{Department of Physics and Astronomy, Texas A\&M University, 4242 TAMU, College Station, TX 77843, USA}
\affiliation{George P. and Cynthia Woods Mitchell Institute for Fundamental Physics \& Astronomy, College Station, TX 77843, USA}

\author[0000-0002-4934-5849]{Daniel Scolnic}
\affiliation{Duke University, Department of Physics, 120 Science Drive, Durham, NC, 27708 USA}

\author[0000-0002-1873-8973]{Benjamin Rose}
\affiliation{Duke University, Department of Physics, 120 Science Drive, Durham, NC, 27708 USA}

\author[0000-0001-5201-8374]{Dillon Brout}
\affiliation{Center for Astrophysics, Harvard \& Smithsonian, 60 Garden Street, Cambridge, MA 02138, USA}
\affiliation{NASA Einstein Fellow}

\begin{abstract}
We compare infrared surface brightness fluctuation (IR~SBF) distances measured in galaxies that have hosted type~Ia supernovae (\snia) to distances estimated from \snia\ light curve fits. We show that the properties of \snia\ found in IR-SBF hosts are very different from those exploding in Cepheid calibrators, therefore, this is a direct test of systematic uncertainties on estimation of the \hubble\ constant (\hnot) using supernovae. The IR~SBF results from \citet{jensen21} provide a large and uniformly measured sample of IR~SBF distances which we directly compare with distances to 25 \snia\ host galaxies. We divide the Hubble flow \snia\ into sub-samples that best match the divergent supernova properties seen in the IR~SBF hosts and Cepheid hosts.  We further divide the \snia\ into a sample with light curve widths and host masses that are congruent to those found in the SBF-calibrated hosts. We refit the light curve stretch and color correlations with luminosity, and use these revised parameters to calibrate the Hubble flow \snia\ with IR~SBF calibrators. Relative to the Hubble flow, the average calibrator distance moduli vary by 0.03~mag depending on the \snia\ subsamples examined and this adds a 1.8\%\ systematic uncertainty to our \hubble\ constant estimate.  Based on the IR~SBF calibrators, $H_0=74.6\pm 0.9$(stat)$\pm 2.7$(syst)~\kmsmpc , which is consistent with the \hubble\ constant derived from \snia\ calibrated from Cepheid variables.  We conclude that IR~SBF provides reliable calibration of \snia\ with a precision comparable to Cepheid calibrators, and with a significant saving in telescope time. 
\end{abstract}


\section{Introduction}

The precise determination of the present expansion rate of the universe, the \hubble\ constant (\hnot), is a fundamental goal of modern cosmology \citep{lemaitre27, hubble29, huchra92, hamuy21}. \hnot\ can be estimated using ``local'' distance indicators, such as type Ia supernovae (\snia), and predicted from the standard $\Lambda$CDM model calibrated with early-universe indicators, such as the cosmic microwave background (CMB). However, the values of \hnot\ determined from local and early-universe indicators are significantly discrepant \citep{riess21, planck20}. This ``tension’’ in the value of \hnot\ could indicate new physics yet to be discovered, or could be a result of unaccounted observational uncertainties \citep[e.g.][]{bernal16,efstathiou21}, or both.

\snia\ are excellent distance indicators for cosmology. Their extreme luminosities allow them to be observable out to high redshifts and their predictable light curves make these supernovae useful as a standardizable candles. To date, \snia\ have been used to constrain the mass density of the universe \citep{perlmutter98, garnavich98}, reveal the cosmic acceleration and dark energy \citep{riess98, perlmutter99}, and they continue to provide a tool for narrowing the uncertainty on \hnot\ \citep{freedman01, riess09, riess19, riess21}.

With significant variations in their peak luminosity, \snia\ are clearly not standard candles. However, empirical correlations between luminosity, color \citep{rpk96}, and light curve shape \citep{phillips93} mean that \snia\ are standardizable candles.  \snia\ properties are also found to correlate with host-galaxy type, with brighter, slower-declining \snia\ typically occurring in spiral galaxies and dimmer, faster-declining \snia\ typically occurring in elliptical galaxies \citep{hamuy95,gallagher05,sullivan06,pruzhinskaya20}. Surprisingly, the mass of a host galaxy is found to correlate with distance residuals after standardization \citep{kelly10, lampeitl10, sullivan10}.

To estimate $H_0$, the luminosity of \snia\ must be calibrated using locally abundant distance indicators, such as Cepheid variable stars, tip-of-the-red-giant-branch (TRBG) stars \citep{freedman19}, or surface brightness fluctuations (SBF) \citep{tonry88, blakeslee09}. In turn, the luminosities of these objects must be calibrated by local techniques and these steps in distance form the basis of the ``distance ladder'' needed to reach the smooth Hubble flow. Much of the current uncertainty in \hnot\ as determined by \snia\ comes from the statistical and systematic errors that accumulate from the lower rungs in the cosmic distance ladder. In particular, Cepheid variable stars have provided a sturdy intermediate step for the \snia\ distance calibration, but testing for their systematic uncertainties is critical to assure the reliability of cosmological studies. 

\snia\ distances can instead be calibrated using TRGB or SBF in an effort to check for the systematic errors in the \hnot\ estimation. SBF is a particularly interesting calibration method as it provides several advantages over Cepheid measurements. SBF does not require resolving individual stars, and can therefore be applied out to about 100~Mpc while maintaining a 5\%\ precision. In comparison, Cepheid calibrators must be within about 40~Mpc to obtain reliable measurements. Further, IR~SBF measurements require only 1 to 2 Hubble Space Telescope ($HST$) orbits per galaxy \citep{jensen21} to achieve a $5\%$ distance precision. This can be compared with the more than $1000$ total $HST$ orbits needed to calibrate 42 \snia\ hosts using Cepheid variables \citep{riess21}. Finally, the SBF method is best applied to elliptical galaxies and therefore targets a complementary population of galaxies and \snia\ than Cepheids, which typically occur in spiral galaxies.

The application of SBF to cosmologically interesting distances was demonstrated by \cite{jensen01} using space-based, near-infrared (IR) imaging. Most recently, HST’s Wide Field Camera 3 (WFC3) has proven to be a superb instrument for SBF distance calibrations, allowing for a median $3.9\%$ distance uncertainty on 63 massive elliptical galaxy distances out to 100 Mpc \citep{jensen21}. 

From IR~SBF distances alone (no \snia), \cite{blakeslee21} provided a robust measurement of \hnot $= 73.3 \pm 0.7$(stat)$\pm 2.5$(syst)~\kmsmpc\ based on a combination of Cepheid and TRGB zero-points. \cite{khetan21} made an estimate of \hnot\ by calibrating \snia\ luminosities from previously published SBF distances compiled from literature sources.  However, their result had significant statistical uncertainties that did not strongly constrain \hnot , and this was primarily due to the inclusion of SBF distances acquired from heterogeneous sources. 

In this paper, we compare the consistent set of IR SBF distances derived from HST imaging of 25 unique \snia\ host galaxies \citep{jensen21} with a consistent set of \snia\ light curve distances from SALT2 \citep{pantheon+1}. Since these SBF distances are calibrated to the Cepheid zero-point, we are still tied to the systematic uncertainties connected to the lower rungs of the distance ladder. Our primary goal is to test for significant systematic differences when calibrating \snia\ in the Hubble flow using IR~SBF distances versus Cepheid distances, given that the two methods target different host-galaxy types. We also test the quality of the IR~SBF distance measurement uncertainties by comparing them directly to the well-determined scatter in the \snia\ luminosities.

The approach taken in this analysis can be considered a ‘Four-Rung Distance Ladder’.  While Three-Rung Distance Ladders have received more attention \citep[e.g.][]{freedman19, riess21}, our approach is able to address possible systematic limitations of the three-rung method.  To summarize, \citet{freedman19} and \citet{riess21} tie geometric distance probes to stellar probes (TRGB and Cepheid respectively), then these stellar probes to supernova probes, and finally supernova probes to redshifts in order to measure the Hubble constant.  Within these three rungs, various choices can be made.  In \citet{riess21}, the \snia\ in the second rung and third rung are just from star-forming galaxies and taken from the Pantheon+ compilation; in \citet{freedman19}, the \snia\ for the second rung are taken from the literature and the third rung is taken solely from the CSP sample.  A separate three-rung approach is done by \citet{blakeslee21}, which use SBF in the role of \snia .  In this approach, either Cepheids or TRGB can be used in the first and second rung.  Our four-rung approach, which allows the usage of a different sample of \snia , is most similar to \citet{khetan21}; this analysis ties geometry to Cepheids, Cepheids to literature SBF, literature SBF to literature \snia , and finally literature \snia\ to more distant literature \snia\ in the Hubble Flow.  In our analysis, we use a single sample of homogeneously calibrated IR~SBF distances \citep{jensen21}, and a homogeneously calibrated set of \snia\ distances \citep{pantheon+2}.  While we use Cepheids in our first rung, \citet{blakeslee21} has shown that combining TRGB and Cepheids changes the resulting value of $H_0$ by only 0.1\%.

A direct comparison of the Cepheid and IR~SBF distance calibrations is limited by the large peculiar velocities found at the calibrator redshifts of $z < 0.02$. Here, we use the large number of Hubble flow \snia\ now available in the Pantheon+ recalibration \citep{pantheon+1, pantheon+2} as a 0.01~mag precision anchor to the nearby host galaxies. We then compare the scatter and the systematics of Cepheid-calibrated \snia\ with those of the SBF-calibrated \snia\ without the need for peculiar velocity models.

A detailed description of our SBF distance and \snia\ light curve data is given in \S~2. Our analysis and comparisons are described in \S~3, followed by our \hnot\ results in \S~4 and discussion in \S~5. A summary of our conclusions is presented in \S~6.

\section{Data}

\begin{deluxetable*}{lccccc}
\centering
%\tablewidth{0.5pt}
\tablecaption{IR SBF Distances to Fornax Hosts \label{ir_sbf}}
\tablehead{
\colhead{Galaxy} & \colhead{$\mu$ (ACS)$^a$} & \colhead{$\mu$ (WFC3/IR)$^b$} & \colhead{difference/$\sigma$} & \colhead{hosted SN} & \colhead{Notes}  \vspace{-0.2cm}
\\
\colhead{ } & \colhead{(mag)} & \colhead{(mag)} & \colhead{ } & \colhead{ } & \colhead{ }
}
\startdata
NGC1380 & 31.609$\pm 0.075$  & 31.465$\pm 0.075$ & 1.36 & 92A &  \\
NGC1404 & 31.503$\pm 0.072$  & 31.453$\pm 0.084$ & 0.45 & 07on, 11iv & \\
NGC1316 & 31.583$\pm 0.065$  & 31.200$\pm 0.093$ & 3.37 & 80N,81D,06dd,06mr & dusty \\
\enddata
\tablenotetext{\tiny a}{Distance moduli from \citet{blakeslee09} corrected by $-0.023$ mag for revised LMC distance.}
\tablenotetext{\tiny b}{IR SBF distance moduli using the analysis described in \citet{jensen21}}

\end{deluxetable*}

\begin{deluxetable*}{lcccccc}
\centering
%\tablewidth{0.5pt}
\tablecaption{Type Ia Supernovae in SBF Galaxies \label{photometry}}
\tablehead{
\colhead{SN} & \colhead{Host} & \colhead{$z$} & \colhead{Stellar Mass} & \colhead{Light Curve} & \colhead{SALT2}  & \colhead{Notes} \vspace{-0.2cm}
\\
\colhead{ } & \colhead{ } & \colhead{(CMB) } & \colhead{(log M$_\odot$)} & \colhead{Source$^b$} & \colhead{Fit$^c$}  & \colhead{ }
}
\startdata
1980N & NGC1316  & 0.00457(03) & 11.77(04) & JRK & yes & Fornax \\
1981D & NGC1316  & 0.00457(03) & 11.77(04) & JRK & large error & Fornax \\
1992A & NGC1380  & 0.00587(15) & 11.22(05) & JRK & yes & Fornax  \\
1995D & NGC2962  & 0.00768(02) & 10.84(05) & CFA1 & yes &   \\
1997E & NGC2258  & 0.01314(12) & 11.59(06) & CFA2 & yes &   \\
1999ej & NGC0495  & 0.01534(03) & 10.93(05) & KAIT, CFA2 & yes &   \\
2000cx & NGC0524  & 0.00697(07) & 11.36(05) & KAIT, CFA2 & color limit &  also 2008Q host  \\
2002cs & NGC6702  & 0.01532(01) & 11.26(05) & KAIT & yes &   \\
2002ha & NGC6964  & 0.01221(10) & 10.96(06) & KAIT, CFA3 & yes &   \\
2003hv & NGC1201  & 0.00511(01) & 10.75(05) & KAIT & yes &   \\
2006dd & NGC1316  & 0.00457(03) & 11.77(04) & JRK & yes &  Fornax  \\
2006ef & NGC0809  & 0.01694(01) & 10.79(07) & CSP, CFA3 & yes &   \\
2006mr & NGC1316  & 0.00551(04) & 11.77(04) & CSP & color limit &  91bg-like; Fornax  \\
2007cv & IC2597  & 0.01362(15) & 11.40(05) & SWIFT & yes &   \\
2007gi & NGC4036  & 0.00505(01) & 9.68(06) & JRK & yes &   \\
2007on & NGC1404  & 0.00594(15) & 11.19(05) & CSP, SWIFT & yes &  also 2011iv host; Fornax  \\
2008L & NGC1259  & 0.01733(05) & 10.85(07) & CFA3 & yes &   \\
2008Q & NGC0524  & 0.00698(02) & 11.36(05) & SWIFT, CFA4 & yes &  also 2000cx host \\
2008R & NGC1200  & 0.01259(15) & 11.37(05) & CSP & yes &   \\
2008hs & NGC0910  & 0.01665(07) & 11.45(06) & SWIFT, CFA4 & yes &   \\
2008hv & NGC2765  & 0.01360(07) & 11.12(05) & CSP, SWIFT, CFA4 & yes &   \\
2008ia & ESO125-G006  & 0.02257(10) & 11.37(10) & CSP & yes &   \\
2010Y & NGC3392  & 0.01142(01) & 10.50(06) & CFA4 & yes &   \\
2011iv & NGC1404  & 0.00519(01) & 11.19(05) & CSP & yes &  also 2007on host; Fornax  \\
%2012fr & NGC1365  & 0.0(01) & none & yes &  Fornax spiral  \\
iPTF13ebh & NGC0890  & 0.01238(05) & 11.36(05) & CSP, SWIFT & yes  &  \\
2014bv & NGC4386  & 0.00578(14) & 10.86(05) & SWIFT & yes  &  \\
2015ar$^a$ & NGC0383  & 0.01600(04) & 11.45(05) & FOUND & yes   &  also 2017hle host\\
2015bo$^a$ & NGC5490  & 0.01697(10) & 11.48(05) & CSP & stretch limit  & 91bg-like  \\
2015bp & NGC5839  & 0.00604(02) & 10.53(05) & SWIFT & yes  &  \\
2016ajf & NGC1278  & 0.01977(05) & 11.37(07) & FOUND & yes  &  \\
2016arc$^a$ & NGC1272  & 0.01218(05) & 11.60(06) & FOUND & bias correction  &  \\
2017hle$^a$ & NGC0383  & 0.01600(08) & 11.45(05) & none & \dots  & also 2015ar host \\
2019ein$^a$ & NGC5353  & 0.00837(01) & 11.37(05) & LOWZ & yes &    \\
%2020ue & NGC4636  & 0.0(01) & none & \dots & &  \\
\enddata
\tablenotetext{\tiny a}{SN2015ar=PSNJ0107203; SN2015bo=SNhunt278; SN2016arc=ASASSN-16ci; SN2017hle=PSP17E; SN2019ein=ATLAS19ieo}
\tablenotetext{\tiny b}{References -- CSP: \citet{burns18}; JRK: \citet{jha07}; KAIT: \citet{KAITM}; SWIFT: \citet{Swift}; CFA1: \citet{CFA1}; CFA2: \citet{CFA2}; CFA3: \citet{CFA3}; CFA4: \citet{CFA4}; FOUND: \citet{FOUND} ; LOWZ: \citet{LOWZ} }
\tablenotetext{\tiny c}{Denotes the success of the SALT2 light curve fit or the reason that the fitter failed.}

\end{deluxetable*}

\subsection{SBF Distances}

Imaging of \snia\ host galaxies was obtained with the Hubble Space Telescope (\hst) using the IR channel of the Wide Field Camera 3 (\wfc) (PropIDs: 11691, PI Goudfrooji; 11712 \& 14219, PI Blakeslee; 14654, PI Milne; 15265, PI Blakeslee). The process of measuring the IR SBF and transforming them to distances is described by \citet{jensen15,jensen21}. 

The IR SBF distances were calibrated with optical SBF distances to 16 Fornax and Virgo cluster galaxies \citep{jensen15,cantiello18}, which are based on the Cepheid zero points \citep{blakeslee09,blakeslee21}. A discussion of the of the SBF calibration to the Cepheid zero-point is given in Appendix~A of \citet{blakeslee10}. The IR~SBF distance moduli have been corrected to the improved LMC distance by shifting $-0.023$~mag \citep{LMC2019}. We use the \citet{blakeslee21} estimate of the zero-point uncertainty in our total error budget (see Section~\ref{sec:hubble}).

IR~SBF distances to Fornax cluster members were not included in \citet{jensen21}. For consistency, the distances from WFC3/IR images (NGC1316, GO~13691; NGC1380 \&\ NGC1404, GO~11712) are estimated here using the same methods as described in \citet{jensen21} and presented in Table~\ref{ir_sbf}. For NGC1380 and NGC1404, the differences between the ACS measurements \citep{blakeslee09} and these WFC3/IR estimates fall within two standard deviations ($2\sigma$) of the combined uncertainties.  

Measurements of NGC1316, the host of multiple \snia , are problematic. The difference in SBF distances between the ACS and WFC3/IR is more than $3\sigma$. The IR~SBF distance to NGC1316 is significantly shorter than the other Fornax galaxies. The origin of this discrepancy is not clear, and we refer the reader to the detailed study by \citet{cantiello13} that analyzes the various distance indicators applied to NGC1316.

It is important to note that the IR~SBF distances used here are tied to the Cepheid zeropoint in the LMC, Virgo, and Fornax clusters. Thus, these results are not a completely independent measurement of the \hubble\ constant. However, early-type hosts and their supernova properties are quite different than supernovae calibrated with Cepheids, and this may result in a systematic \hnot\ offset between the two methods when extended out to the Hubble flow. To test for systematic offsets, we compared the calibrator samples with a large set of Hubble Flow \snia\ to search for a magnitude shift that is sample dependent.  An IR~SBF calibration of SNe~Ia that is completely independent of Cepheids will require further direct geometrical and TRGB distances to nearby elliptical galaxies out to the distance of the Virgo and Fornax clusters.

\subsection{Supernova Light Curves}

Empirically, \snia\ luminosities correlate with the rate of fading after maximum brightness \citep{phillips93} and their color near maximum \citep{rpk96}. Host galaxy properties, such as total stellar mass, also show some correlation with \snia\ brightness after correcting for light curve shape and peak color  \citep{kelly10}.

There have been several light curve ``fitters'' developed since the original luminosity-light curve shape relation was identified including $\Delta m_{15}(B)$ \citep{hamuy96}, MLCS \citep{rpk96}, MLCS2k2 \citep{jha07}, SALT \citep{guy05}, and SNooPy \citep{burns11}. The Spectral Adaptive Light Curve Template-2 \citep[SALT2;][]{guy07} is an improvement on the original SALT algorithm and has been recently revised \citep[e.g.][]{taylor21}.

SALT2 is an empirical model of the spectral evolution of \snia . The model is trained using a large set of supernova spectra and light curves that define a ``surface'' in wavelength, phase and, flux \citep{mosher14}. The surface is a time-varying spectral energy distribution (SED) that is convolved with redshifted filter transmission functions to generate photometric models. The model photometry is used to fit observed \snia\ light curves and extract stretch $x1$, color $c$, and peak magnitude, $m_B$ estimates. These parameters represent the best fit model of the rest-frame light curve observed in a standard $B$ filter. \snia\ luminosities are corrected to a fiducial light curve through a stretch correlation, $\alpha$, and a color correlation, $\beta$ \citep{tripp98}.  This form of SNIa distance estimation is often called the Tripp Relation.  The \snia\ distance modulus, $\mu$, is then
\begin{equation} \label{tripp}
\mu\; =\; m_B+\alpha\;x1-\beta\;c - M_B - \delta_{bias}
\end{equation}
where $m_B$ is a parameter that represents the observed peak magnitude of the \snia\ and $M_B$ is the absolute peak magnitude of the fiducial light curve defined at $x1=0$ and $c=0$. The final term, $\delta_{bias}$, is a multi-dimensional bias correction based on SNANA simulations and it is described by \citet{pantheon+1}.

We used the publicly available \snia\ light curve data described by \citet{pantheon+1} and \citet[][e.g. Pantheon+]{pantheon+2} that employed a retrained SALT2 algorithm. Distances are based on bias corrections estimated using simulations implemented through the SNANA software package \citep{kessler09}. 

When stepping up the distance ladder,  \hnot\ is estimated by calibrating $M_B$, the peak absolute magnitude of a bias-corrected supernova with normal stretch and color. This is done by estimating the distance to nearby \snia\ host galaxies using local distance indicators such as Cepheids, TRGB, SBF, and others. The average value of $M_B$ derived from these calibrators is then applied to \snia\ out in the Hubble flow where peculiar velocities are small compared with the expansion velocity. A difference in this paper, is that we work with distance moduli directly. This is convenient because the SALT2 fitter delivers a distance modulus, $\mu$, for each light curve and the calibrator distances are provided as distance moduli \citep[e.g.][]{jensen21}. Consequences of this parameterization are discussed in Section~\ref{subsets}.

We selected a subset of the Pantheon+ light curves with $0.02 < z_{\rm CMB} < 0.25$ to use as the ``Hubble Flow'' sample. At redshifts less than 0.02, the distance uncertainties are dominated by galaxy peculiar velocities. The high redshift cut was chosen to avoid a mismatch between the Hubble flow sample and the supernovae typically found in SBF hosts. As we shall see, SNIa in SBF hosts tend to be less luminous than those in actively star-forming galaxies so less likely to be identified in high redshift searches.

\subsection{Light Curves for \snia\ in SBF Hosts}

Table~\ref{photometry} lists 33 \snia\ events that were observed in galaxies with IR~SBF distances measured by \citet{jensen21} or given in Table~\ref{ir_sbf}. 

\snia\ occurring in elliptical galaxies tend to have fast declining light curves and there is overlap between the normal \snia\ and SN~91bg-like events. SN~91bg-like events are considered peculiar due to their intrinsically red color at maximum light and strong Ti~II absorption in their spectra \citep[e.g.][]{garnavich04}. Our SBF sample excludes \snia\ that were classified as SN~91bg-like from their spectra. The SNANA simulations further removed supernovae that exceed color, stretch, or bias limits (column labeled ``SALT2 Fit''). There was no published light curve data available for SN~2017hle. Our sample contains 27 \snia\ with reliable light curves fits after making these cuts.

For several supernovae there are light curves derived from multiple sources. The Pantheon+ sample calculates separate SALT2 fits for each light curve source. When a supernova had multiple fits, we averaged the distance and SALT2 parameters weighted by their uncertainties. Here, the average difference in SALT2 distance moduli between light curves obtained for the same supernova is 0.027~mag. 

``Sibling'' supernovae are multiple events that occurred in the same host galaxy \citep{scolnic20}. In our sample,  there were several IR~SBF galaxies that hosted two or more supernovae. However, after light curve bias cuts, only NGC~1404 (SN~2007on and SN~2011iv) and NGC~1316 (SN~1980N and SN~2006dd) had surviving siblings. Given the small number of siblings, we treat them as independent calibrators.

\subsection{Light Curves for \snia\ in Cepheid Hosts}

To compare the IR-SBF distances and their associated supernovae with a set of Cepheid-calibrated hosts and their supernovae, we use the ``SH0ES'' \citet{riess16} sample of \snia . This set of 19 hosts with Cepheid calibrated distances have been recently updated in the \citet{riess21} compilation.  As a test of consistency, these Cepheids are used to calibrate the same set of Hubble flow supernovae as is done with the IR~SBF distances.

\section{Analysis}
\subsection{Direct Comparison of SALT2 and SBF \snia\ Distances}

\begin{figure}
    \centering
    \includegraphics[width=\columnwidth]{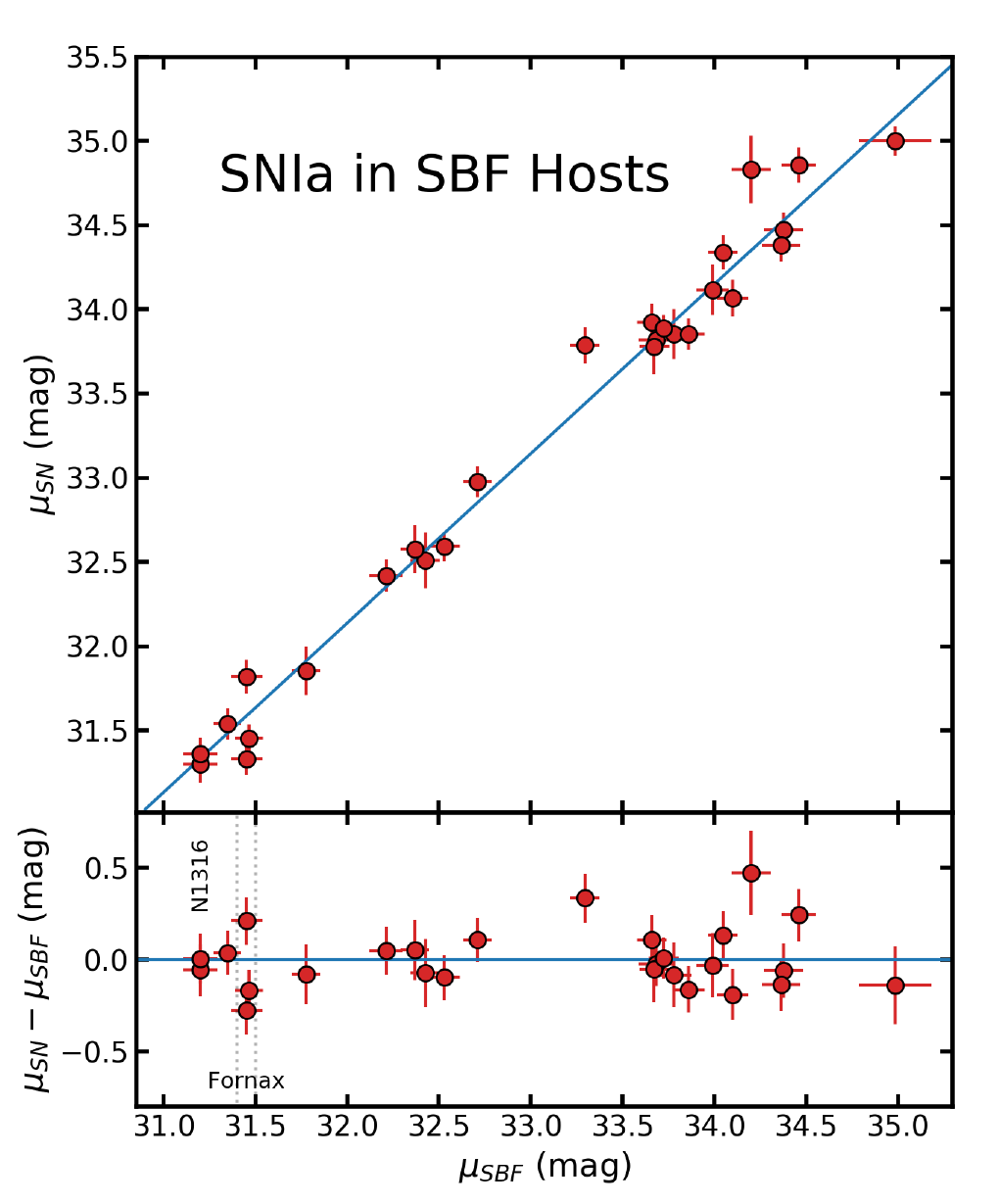}
    \caption{{\bf top:} The Pantheon+ light curve distances versus the host galaxy IR~SBF distances. Distances for \snia\ with multiple sources of photometric data have been averaged.   {\bf bottom:} The difference between the \snia\ and SBF distance moduli as a function of the SBF distance. The average difference in distance moduli has been subtracted. }
    \label{sn_sbf}
\end{figure}

We compare the SBF distances \citep{jensen21} with the SALT2 \snia\ distances in Figure~\ref{sn_sbf} (top), where \snia\ distance modulus uncertainties are taken from the Pantheon+ fits. Because we are directly comparing distance moduli without use of redshifts, we do not add uncertainties due to peculiar velocities. The differences between distance moduli are also shown in Figure~\ref{sn_sbf} (bottom), where their uncertainties have been combined in quadrature.

%*********************

Assuming the \snia\ distances and the IR SBF distances differ by a constant, a $\chi^2$ minimization provides an offset of $\Delta M=0.156\pm 0.027$ mag in distance modulus. The $\chi^2$ value for the single parameter offset model is 32.4 for 26 degrees of freedom, indicating that the scatter is in fair agreement with the combined uncertainties.

Looking at Figure~\ref{sn_sbf} it is clear that a significant amount of scatter comes from a single Fornax cluster galaxy, NGC~1404, that hosted two \snia\ \citep{ashall18,gall18}. Because SN~2007on and SN~2011iv are essentially at the same distance, we can directly compare the difference in their SALT2 distance moduli and their uncertainties. Their distances differ by nearly 0.5~mag (3.5$\sigma$), which should occur at random for only 1 in 2000 siblings \citep{scolnic20}.  Recalculating the scatter without including the two events in NGC~1404, gives $\chi^2=$25.2 for 24 degrees of freedom, suggesting that the  combined IR~SBF and \snia\ error estimates for the remaining sample are consistent with the observed scatter. As we have no objective reason to exclude NGC1404 or its supernovae, they will remain in the sample for the rest of the analysis.

NGC~1316 is another Fornax cluster galaxy with issues and a detailed review is given by \citet{cantiello13}. It has hosted multiple \snia\ and two (SN~1980N and SN~2006dd) made it through all the light curve cuts. However, its IR~SBF distance (Table~\ref{ir_sbf}) is 0.38~mag (3.4$\sigma$) shorter than the \citet{blakeslee09} SBF distance. The new IR~SBF distance (Table~\ref{ir_sbf}) places it in the foreground of the Fornax cluster. NGC~1316 is a giant elliptical galaxy, but with a dusty core that may impact the SBF estimates as well as the \snia\ distances. SN~1980N occurred far from the galaxy center, but SN~2006dd was projected on a dusty region near the core. Surprisingly, the galaxy does not standout in Figure~\ref{sn_sbf} because both \snia\ distances are in good agreement with its IR~SBF distance. 

As the IR~SBF distances to sibling supernovae are not independent in our calculation, we can also average their SALT2 distances and errors to recalculate the $\chi^2$ parameter. After averaging the \snia\ distances to NGC~1404 and NGC~1316 we see that $\chi^2=$25.2 for 24 degrees of freedom.

An important point is that the \snia\ distance uncertainties are empirically well determined from the large set of Hubble flow events where the scatter added by peculiar velocities is small. That the reduced $\chi^2_\nu \approx 1$ for our analysis demonstrates that the uncertainties given in \citet{jensen21} accurately describe the scatter in applying the IR~SBF method. That the IR~SBF uncertainties are typically 0.09~mag implies that the precision of the IR~SBF calibration to an individual galaxy is comparable to that of Cepheid calibrators. 

\subsection{Comparison with Khetan et al.}

\citet{khetan21} used SBF distances from a wide variety of heterogeneous sources to make an estimate of $H_0$. Here, we use the IR SBF distances from \citet{jensen21} that have been uniformly analyzed. Only seven of our SBF hosts are included in the \citet{khetan21} sample and for these galaxies, the average difference in distance modulus is 0.17$\pm 0.11$~mag.

The difference between the heterogeneous SBF distances taken from the literature and the consistent sample analyzed here can be seen in the intrinsic scatter of the calibrators.  \citet[][Table~4]{khetan21} shows an
additional scatter in the calibrator brightnesses of 0.29~mag. In contrast, the scatter we measure using IR~SBF distances from \citet{jensen21} is consistent with an uncertainty of 0.09~mag in distance modulus. 

%**********************
\begin{figure}
    \centering
    \includegraphics[width=\columnwidth]{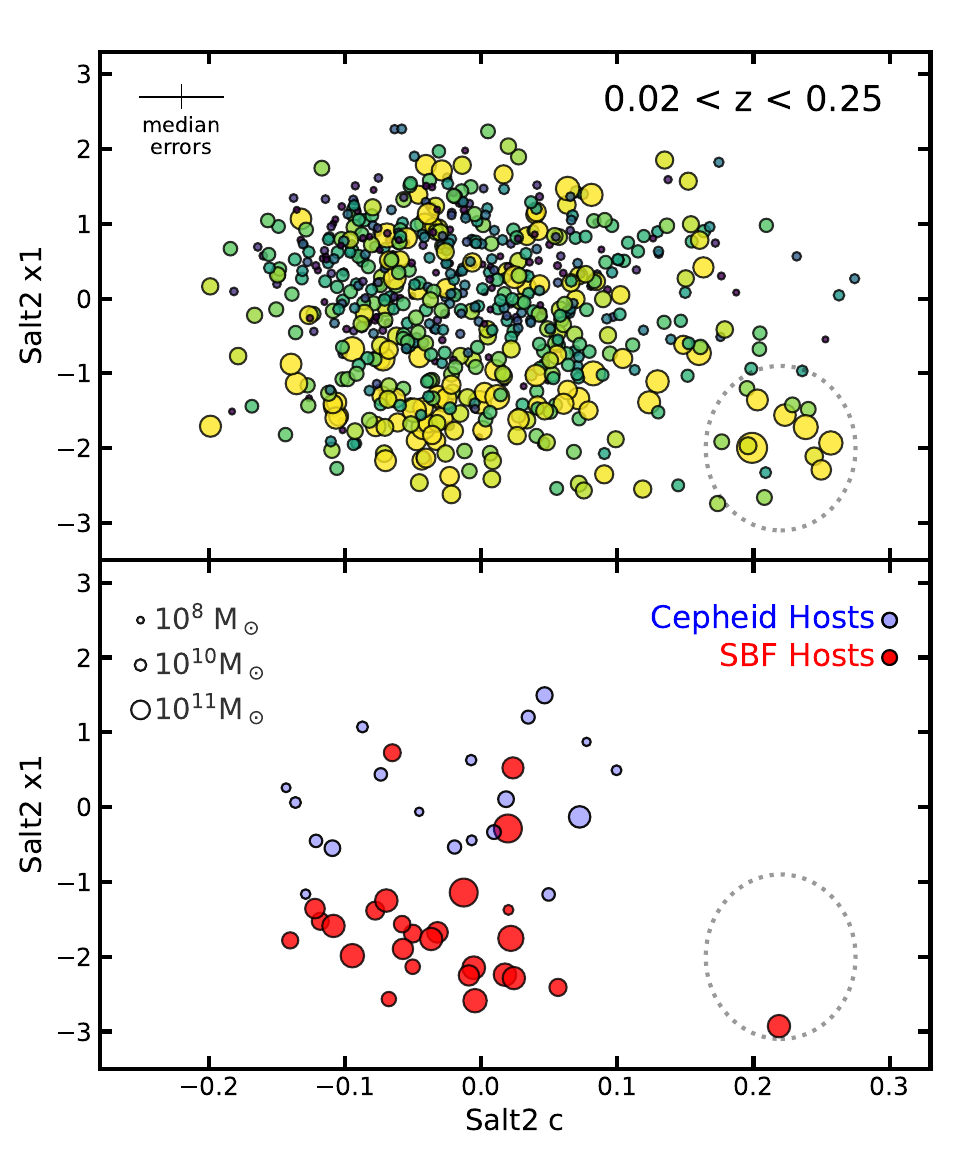}
    \caption{{\bf top:} The Hubble flow sample showing the SALT2 stretch parameter, $x1$, versus the SALT2 color parameter, $c$. The the point size and color indicate the relative stellar mass of the host galaxy (yellow for more massive hosts). While a wide range of host masses are seen for events with $x1>-1$, fast-declining \snia\ with $x1<-1$ tend to occur in very massive hosts. A group of fast declining, red ($c\gtrsim 0.2$) supernovae appear in the lower-right corner of the plot as indicated by the dotted circle. {\bf bottom:} Same as the top plot, but showing the SBF-calibration sample (red) and Cepheid-calibration sample (blue). The SALT2 color distributions are similar between all three samples, but the SBF-calibrated \snia\ tend to exhibit fast declining light curves ($x1<-1.0$) while Cepheid-calibrated \snia\ exhibit slower decline rates ($-1\lesssim x1 \lesssim 2$). One SBF supernova,  SN2016ajf, lands in the lower-right of the plot and appears to be a member of the grouping of points noted in the Hubble flow sample.}
    \label{color_stretch}
\end{figure}
% turn color bar for lower plot into a legend
%***********************

\subsection{Comparison of \snia\ in SBF- vs. Cepheid-Calibrated Hosts}

\subsubsection{Light Curve Stretch}

\begin{figure}
    \centering
    \includegraphics[width=\columnwidth]{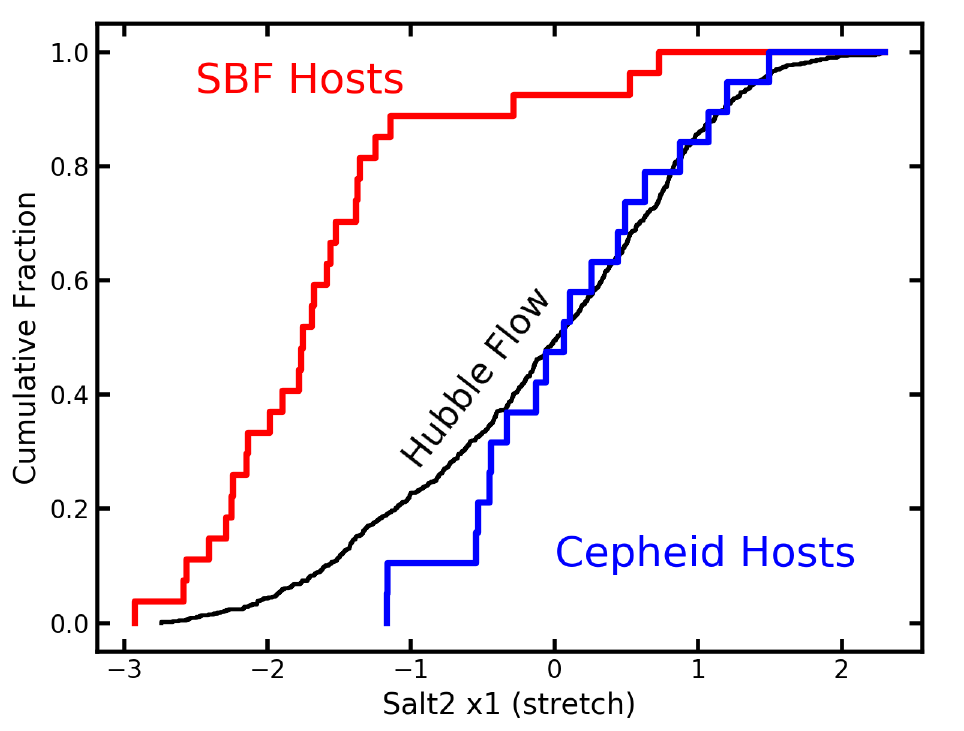}
    \caption{The cumulative distribution of the SALT2 light curve width parameter $x1$ for the SBF-calibration sample (red), the Cepheid-calibration sample (blue), and the Hubble flow sample (black). $89\%$ of the SBF-calibrated \snia\ have $x1<-1.0$ and $89\%$ of the Cepheid-calibrated \snia\ have $x1>-1.0$.
    A minority ($22\%$) of the Hubble flow \snia\ are fast-decliners with $x1<-1.0$.}
    \label{cumulative}
\end{figure}

The IR~SBF method is best applied to massive elliptical galaxies \citep{blakeslee21} with old stellar populations. SBF-calibrated SNIa hosts provide a stark contrast to the Cepheid-calibrated hosts. Cepheid variables are exclusively found in late-type galaxies with young stellar populations. Recent star formation is a prerequisite for containing massive stars that are caught evolving through the instability strip. In addition, \snia\ in passive galaxies tend to have faster-declining light curves (low stretch values) than \snia\ in star-forming hosts \citep{hamuy96,gallagher05,sullivan06}.
Figure~\ref{color_stretch} compares the SALT2 parameters between the Hubble flow \snia, the SBF-calibrated sample, and the Cepheid-calibrated sample. The SALT2 stretch parameter in the SBF-calibrated sample are seen to be significantly smaller than in the Cepheid-calibrated sample. The transition in light curve widths for \snia\ in the SBF-calibrated hosts and Cepheid-calibrated hosts is extremely sharp. The cumulative distributions for the two samples (Figure~\ref{cumulative}) shows that $89\%$ of the \snia\ in SBF-calibrated hosts have $x1<-1$ while the same fraction of the \snia\ in Cepheid-calibrated hosts have $x1>-1.0$. Therefore, the stretch parameter of $x1=-1$ is a natural place to divide the Hubble flow sample when calibrating distances between the two methods.

\subsubsection{Color at Peak}

Figure~\ref{color_stretch} shows that the distribution of the SALT2 color parameter, $c$, is similar between the Cepheid-calibrated sample and the SBF-calibrated sample. Based on spectroscopic classifications, we have avoided using SN1991bg-like \snia\ in the SBF sample. Their colors tend to be redder than normal \snia\ at maximum and their use as standardizable candles is problematic \citep{garnavich04}. 
We note that in the Hubble Flow sample, there are a number of fast decliners ($x1\approx -2$) with reddish peak colors ($c\approx 0.2$) that form a distinct grouping in the lower-right of the Hubble flow sample shown in Figure~\ref{color_stretch}. These passed the color and bias cuts from the SNANA simulations, but are significantly more red than a majority of the SBF-calibrated \snia. One supernova in the SBF-calibration sample (SN2016ajf) shares the peak color and stretch of the group. This small subset of events may represent a transition between normal \snia\ and the peculiar SN1991bg-like class. Further analysis of these ``narrow-normal'' \snia\ is addressed in Milne et al. (in prep). %(make comment about narrow-normal upcoming paper)

%**********************
%***********************

%**********************
\begin{figure}
    \centering
    \includegraphics[width=\columnwidth]{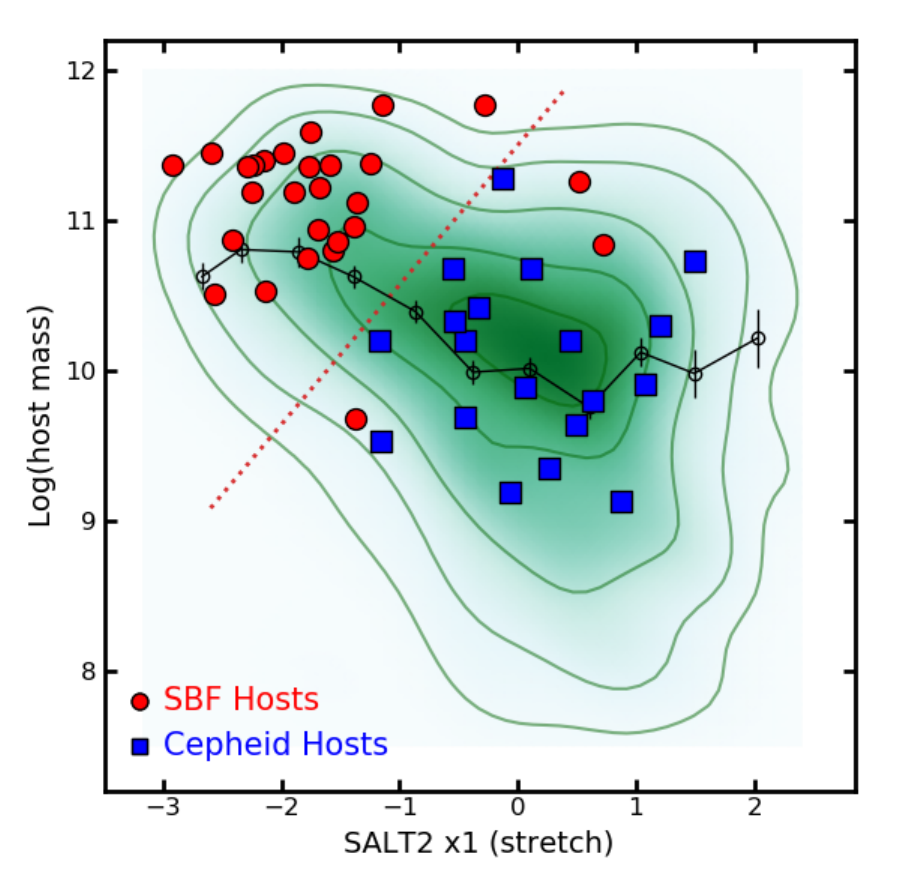}
    \caption{Host stellar mass versus the SALT2 stretch parameter for SBF-calibrated \snia\ (red) and Cepheid-calibrated \snia\  (blue). The Hubble flow sample is shown as density contours and as color intensities using a Gaussian density estimator. The open circles show the median log host stellar mass binned in $x1$. For $x1<-1$, the median host mass is 10$^{10.8}$~M$_\odot$, while for $x1 > -1$ the median host mass remains close to 10$^{10.0}$~M$_\odot$. For further analysis, we adopt the stretch/mass division between ``SBF-like'' \snia\ and ``Cepheid-like'' \snia\ shown as the diagonal dotted line (see Section~\ref{subsets}).}
    \label{mass}
\end{figure}
%***********************

\subsubsection{Host Masses}

The stellar masses of the SBF-calibrated host galaxies were estimated using the prescription in \citet{ma14}. The 2MASS $K_{\rm m, ext}$ magnitude is converted to an absolute magnitude using the distances from \citet{jensen21} after correcting for $K$-band Milky Way dust extinction. The stellar mass estimates are given in Table~\ref{photometry}. For the Cepheid-calibrated hosts and the Hubble flow hosts, we take the logarithm of the stellar mass from the Pantheon+ release \citep{pantheon+2}.
% $K$\_$m$\_$ext$

%**********************
\begin{figure*}
    \centering
    \includegraphics[scale=0.56]{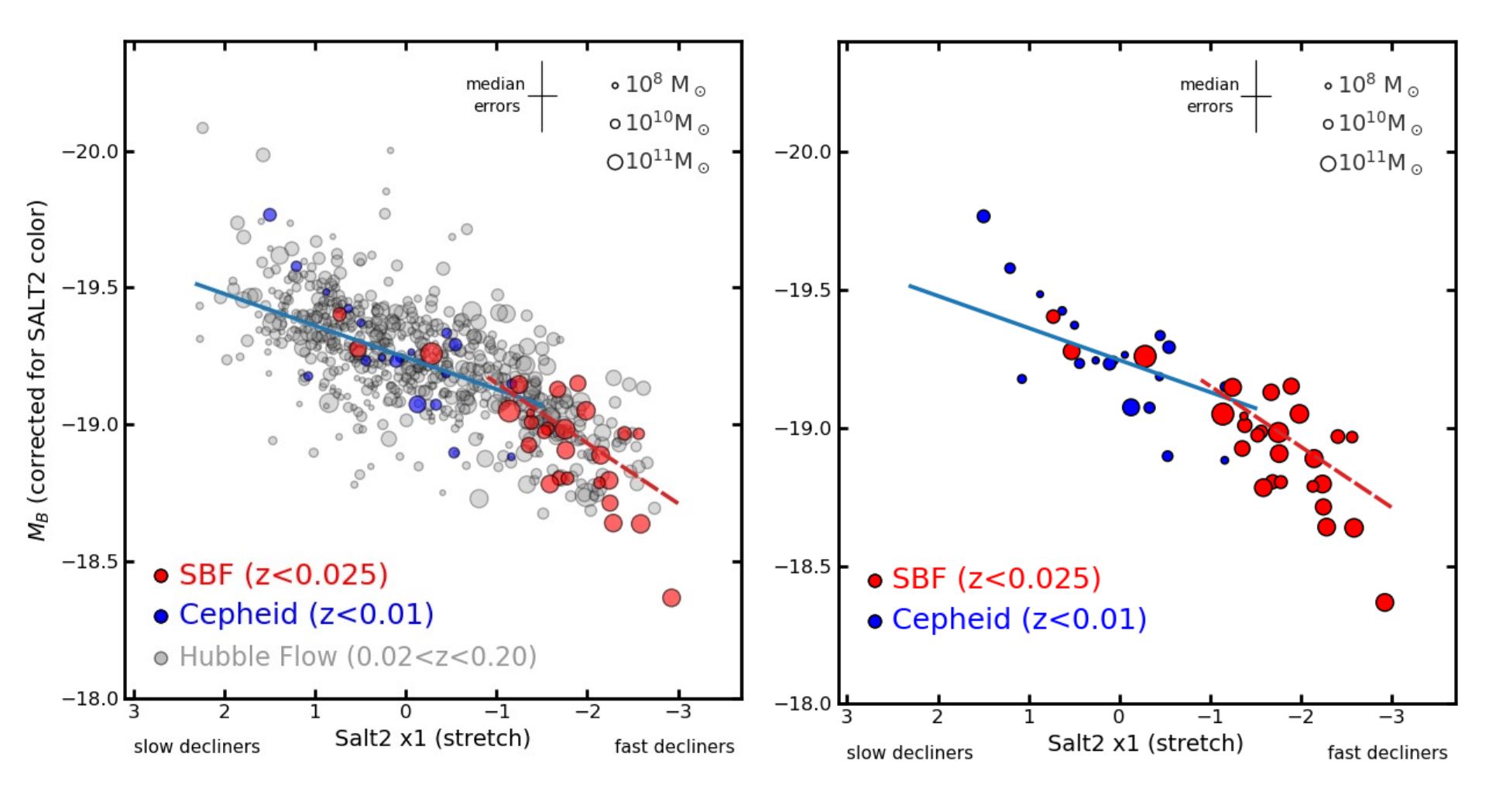}
    \caption{The \snia\ peak luminosity versus SALT2 stretch parameter (a.k.a. the Phillips diagram)  showing the \snia\ in SBF calibrators (red), Cepheid calibrators (blue) and, in the Hubble flow (grey). For clarity, the right panel is a copy of the left panel without the Hubble flow supernovae plotted. For the Hubble flow sample (grey points), the \snia\ peak luminosities have been corrected for SALT2 color and distances calculated from the CMB frame redshifts with an assumed $H_0=73$~\kmsmpc. No Pantheon+ bias corrections have been included in the plotted absolute magnitudes.  SBF distances and Cepheid distances are used to calculate luminosities of the SBF-calibrated (red) and Cepheid-calibrated (blue) \snia\ respectively. The \snia\ host-galaxy mass is indicated by the size of the point. The best fit relation between $M_B$ and $x1$ for events with $x1>-1.0$ is shown as the solid blue line.  The dashed red line is the same but for $x1<-1.0$. The difference suggests that luminosity correction for SALT2 stretch is non-linear.
    \label{phillips} }
\end{figure*}
%***********************

The distribution of host masses shown in Figure~\ref{color_stretch} suggests that the SBF-calibration sample has significantly more massive hosts than seen in the Cepheid-calibration sample. There are several factors that may contribute to this. For example, SBF is typically applied to bright (thus massive) elliptical galaxies whether or not they have hosted \snia. Thus, a few \snia\ in our sample exploded in ellipticals with pre-existing SBF measurements. SN2019ein is an example as its host, NGC~5353, was observed as part of the MASSIVE survey \citep{ma14}.

Another factor favoring massive hosts in the SBF-calibration sample is that the rate of \snia\ explosions in passive galaxies appears to be proportional to the host stellar mass, while in star-forming hosts the \snia\ rate is dominated by the current star formation rate \citep{scannapieco05,sullivan06}. So, for elliptical galaxies with no (or very little) star formation, \snia\ are most likely to appear in hosts with high stellar masses.

%**********************
\begin{figure*}
    \centering
    \includegraphics[scale=0.39]{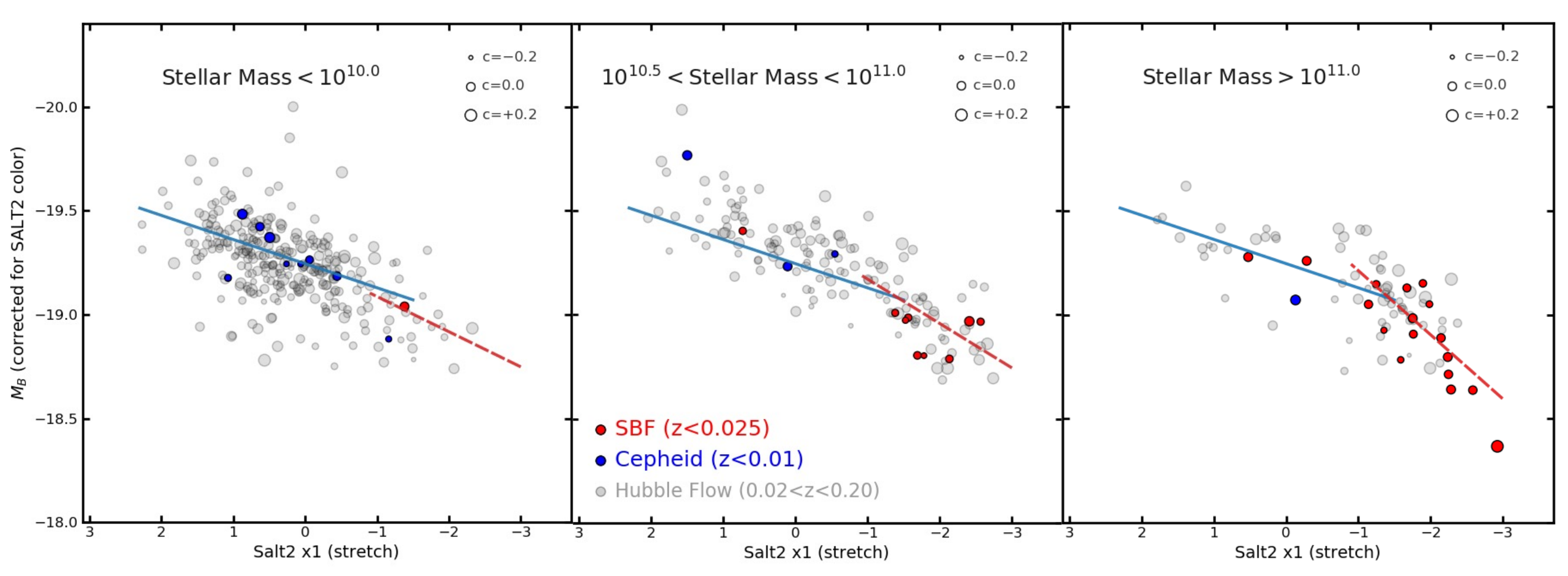}
    \caption{The Phillips relation divided into host stellar mass bins. In this figure, the sizes of the circles represent the SALT2 color of the SN Ia at peak. No Pantheon+ bias corrections have been included in the plotted absolute magnitudes. The solid blue line in all the panels shows the best fit to all \snia\ with $x1>-1$ as calculated for Figure~\ref{phillips}. The blue points show the \snia\ in Cepheid-calibrated hosts and the red points are \snia\ in SBF-calibrated hosts. {\bf left:} The Phillips relation for \snia\ in low-mass hosts. The red dashed line is a fit to the points with $x1<-1$ and the slope is similar to that of the slow decliners. {\bf center:} The Phillips relation again, but for intermediate mass hosts. Note that \snia\ with $x1>-1$ generally sit above the blue line indicating the ``mass-step'' has been crossed. The shift in luminosity between the left and center panels for $x1>-1$ is 0.10~mag. For \snia\ with $x1<-1$, the slope has steepened slightly from the left panel at $0.16$ to $0.21$ for the intermediate masses. {\bf right:} The Phillips relation for very high mass hosts. Note that there are very few slow decliners for this mass range. For $x1<-1$, the best fit slope further steepens to $\alpha = 0.31$.
    \label{phillips_mass} }
\end{figure*}
%***********************

The distinction between \snia\ in SBF-calibrated hosts and Cepheid-calibrated hosts is well shown in Figure~\ref{mass}, where the SALT2 stretch parameter is plotted against host stellar mass. For fast-fading \snia\ ($x1<-1$) in the Hubble flow, the median host mass is 10$^{10.8\pm 0.10}$~M$_\odot$ , while the median mass drops by nearly an order of magnitude to 10$^{9.93\pm 0.08}$~M$_\odot$ for galaxies hosting slowly fading supernovae. The vast majority of SBF-calibrated host galaxies are significantly more massive than Cepheid-calibrated host galaxies.

A relation between residuals on the \hubble\ diagram and the stellar mass of \snia\ host galaxies was noted by \citet{kelly10} and confirmed by \citet{sullivan10} and \citet{lampeitl10}. An empirical correction has commonly been applied as a ``mass step'' \citep[e.g.][]{betoule14,kelsey21} occurring at $M_\star=10^{10}$~M$_\odot$, although the physical origin of the correlation remains uncertain. Parameters such as host mass, metallicity, stellar age, star-formation rate, and light curve width show correlations that may be related to trends in galactic evolution and the \snia\ delay-time distribution \citep{rose19,childress14}. Recently, \citet{brout21} argued that the apparent mass step is a result of differing dust properties between massive and low-mass hosts and it is not intrinsic to \snia\ luminosities. 
The bias corrections applied to the Pantheon+ light curves mitigates the mass-step issue in the resulting \snia\ distance estimates. 

\subsubsection{Phillips Relation}

The relation between light curve decline rate and peak luminosity for \snia\ is often referred to as the Phillips relation \citep{phillips93}. We plot the Phillips relation for the Hubble flow, the SBF-calibrated, and Cepheid-calibrated samples in Figure~\ref{phillips}. The SALT2 stretch parameter, $x1$, is used as the light curve decline rate indicator. The luminosity is estimated from the apparent peak magnitude, $m_B$, corrected for the SALT2 color term using $\beta=3.09$. No correction for stretch has been applied in Figure~\ref{phillips}. For the Cepheid and SBF-calibrated \snia, we adopted distances from \citet{riess21} and \citet{jensen21}. For the purposes of plotting the absolute magnitudes that are consistent with the calibrators, we calculated the distance moduli for the Hubble flow sample from their redshift in the CMB frame and corrected for cosmological parameters assuming \hnot=73~\kmsmpc\ . Here, the choice of \hnot\ is arbitrary and this value was chosen to approximate the expected supernova absolute magnitudes. Selecting \hnot=68~\kmsmpc\ would shift all the points by 0.15~mag along the y-axis. No Pantheon+ bias corrections have been applied to the supernovae shown in Figure~\ref{phillips}. 

A weighted linear regression for all \snia\ with $x1>-1$ gives a slope of $\alpha_{slow}=0.11\pm0.02$ while the slope for the fast decliners with $x1<-1$ is $\alpha_{fast}=0.212\pm0.023$. Removing the extremely fast and red SN2016ajf from the fit, changes the estimated slope to $-0.207\pm 0.023$ (a shift of 0.2$\sigma$).
The $\alpha$ calculated from the entire Pantheon+ bias-corrected set \citep{pantheon+2} gives a slope of $\alpha =0.148\pm 0.003$, which lies between the slopes found for the sets divided by stretch. 

%**********************
\begin{figure}[h!]
    \centering
    \includegraphics[scale=0.5]{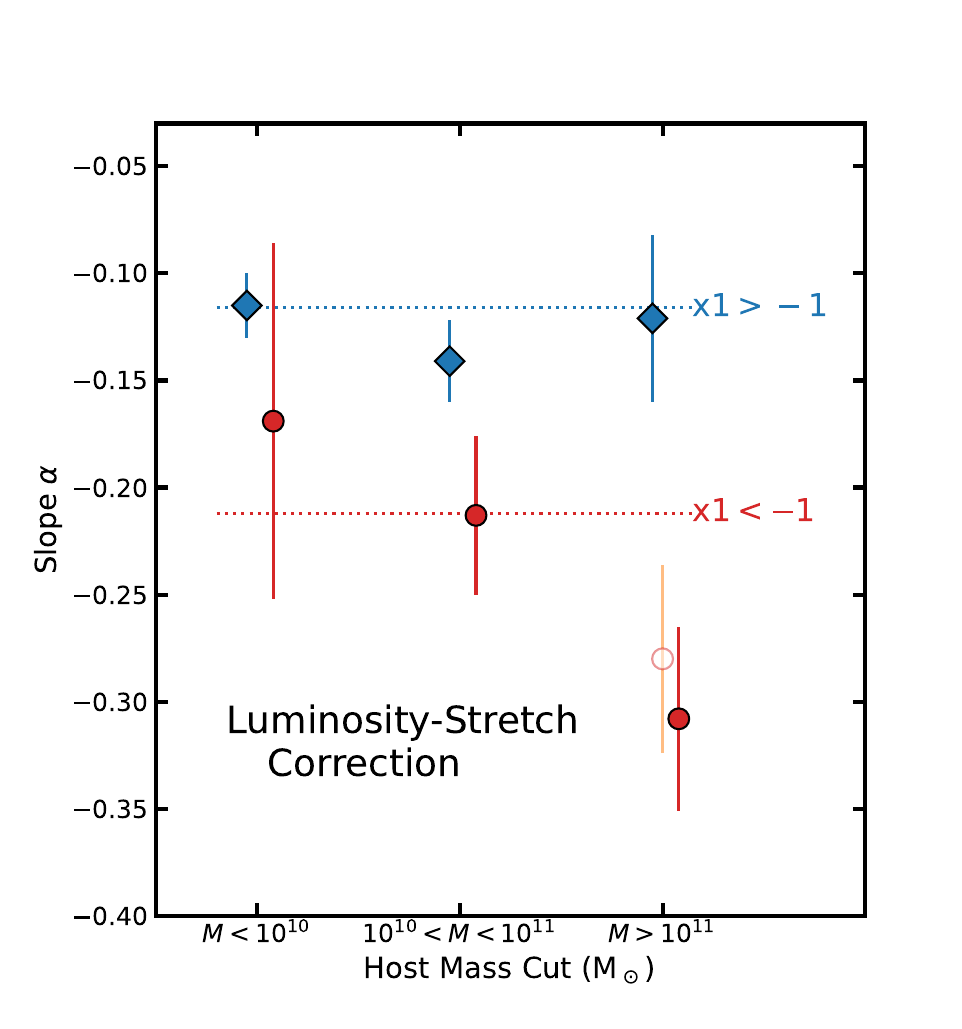}
    \caption{The slope, $\alpha$, of the correlation between luminosity and light curve stretch for various subsets of \snia\ discussed in Section~3. The \snia\ were divided into slow decliners with stretch parameters of $x1>-1$ (shown as blue diamonds), and fast decliners with stretch parameters of $x1<-1$ (shown as red circles). The dotted lines show the slope estimated using the full range of host galaxy masses. The points indicate the slopes measured using \snia\ divided into three host mass bins. The open circle shows the measured slope when the extreme SN2016ajf is excluded from the fit in the highest host mass bin.}
    \label{slopes}
\end{figure}
%***********************

Without the bias-correction, the stretch/luminosity correction in the Tripp relation (Equation~\ref{tripp}) should include a non-linear term to account for the sharp change in slope between fast and slow fading \snia . The Phillips relation using the $\Delta m_{15}(B)$ was improved by applying a quadratic function \citep{phillips99}. Recently, \citet{burns18} applied a color-stretch parameter that shows a clear non-linear relationship between light curve decline rate and luminosity ranging from ultraviolet to the near-infrared bands.  \citet{ashall20} found a change in the color-stretch parameter luminosity relation near $s_{BV}=0.71$ which corresponds to a SALT2 stretch of $x1\approx -1.5$. This transition corresponds to an absolute magnitude of $M_B=-19.0$

Because the SBF- and Cepheid-calibration samples radically differ in host stellar mass, we show the Phillips relation for three mass cuts in Figure~\ref{phillips_mass}. We again see that the Cepheid-calibrated galaxies dominate the low-mass bin while the SBF-calibrated galaxies are found mainly in the intermediate-mass and high-mass sub-samples. Again, the Pantheon+ bias corrections have not been applied for these plots.

For the slow declining \snia\ ($x1>-1.0$), there is very little change in the luminosity slope for the different mass cuts, but the mass-step is clearly seen going from the low-mass to intermediate-mass samples. That is,  the \snia\ in intermediate-mass hosts tend to be nearly 0.1~mag brighter than \snia\ in low-mass hosts. Given the relatively few slow-decliners in the very highest mass bin, it is difficult to define a trend there.

For fast-decliners ($x1<-1.0$), the error-weighted linear regression used to estimate the slope in Figure~\ref{phillips_mass} steepens with increasing host mass. For the highest host-mass bin, the luminosity slope is twice the value measured for the full sample, that is, $-0.308\pm 0.043$ versus $-0.135\pm 0.005$ (a difference of $4.0\sigma$). When SN2016ajf is excluded from the fit, the slope of the highest host-mass sample becomes $-0.280\pm 0.045$ (a shift of $0.7\sigma$).

In Figure~\ref{slopes}, we summarize the estimated slopes (and uncertainties) for the fast and slow decliners over three host mass ranges. For light curves with $x1>-1$, the estimated slopes are consistent over all three host mass bins. For fast declining \snia\ ($x1<-1$), there is an indication that $\alpha$ gets steeper with increasing host mass.

Figures~\ref{phillips} and \ref{phillips_mass} suggest that it is risky to apply a simple Tripp relation to \snia\ found in massive, early-type hosts such as those used in measuring SBF distances. The bias corrections calculated for the Pantheon+ distances \citep{pantheon+2} attempt to mitigate the observed non-linear variations over stretch. A complementary approach used here is to isolate Hubble flow supernovae that share characteristics of \snia\ in SBF-calibrators and to refit the linear Tripp parameters.   

\subsection{The \hubble\ Constant } \label{subsets}

The previous sections show that \snia\ events in Cepheid-calibrated hosts and SBF-calibrated hosts are very different in their properties.  While their colors near maximum have similar distributions (see Figure~\ref{color_stretch}), \snia\ in Cepheid-calibrated hosts tend to be slow decliners in low-mass hosts while \snia\ in SBF-calibrated hosts tend to be fast decliners in high-mass hosts.  To understand the impact on \hnot\ that may result from calibrating Hubble flow with supernovae and hosts that possess strongly divergent properties, we will first estimate \hnot\ based on the full Hubble flow sample and then by dividing the Hubble flow sample in to subsets that match the properties of the calibrators. 

The ``full'' Hubble flow sample consists of 816 \snia\ with $0.02 < z_{\rm CMB} < 0.25$. The full sample is divided into ``fast'' and ``slow'' sub-samples at $x1=-1.0$. To test for systematics, the fast sub-sample is calibrated with the \citet{jensen21} SBF distances and the slow sub-sample is calibrated using the Cepheid distances from SH0ES \citep{riess16,riess21}. The \snia\ distances in the ``full'', ``fast'', and ``slow'' sub-samples are taken from the bias-corrected  Pantheon+ sample \citep{pantheon+2}

The SBF calibration sample consists of supernovae and their host galaxies that are distinct from the typical \snia\ in the Pantheon+ sample. The SBF-calibration sample is dominated by fast-declining supernovae that have exploded in massive, passive galaxies. To select a Hubble flow sample as close as possible to the characteristics of events in the SBF-calibration sample, we must use \underline{both} the light curve stretch parameter and the host mass to define a ``SBF-like'' sample. This is an odd name, but it concisely captures the goal of matching the characteristics of the SBF-calibrated galaxies and their supernovae. We define the SBF-like sub-sample in both stretch and host mass using the relation, 
\begin{equation}
\log_{10}({\rm mass}) = 0.93(x1) + 11.5\;,
\end{equation}
as shown by the red dotted line in Figure~\ref{mass}. Events above this relation
are designated ``SBF-like.'' The SBF-like sample consists of 162 Hubble flow supernova and 24 SBF-calibrators. For this sub-sample we do not use the Pantheon+ bias corrections. Instead, selecting only the SBF-like events, we refit the SALT2 stretch and color coefficients to minimize scatter in the Hubble flow and apply these revised coefficients to the \snia\ hosted by SBF-calibrated galaxies.

%**********************
\begin{figure*}
    \centering
    \includegraphics[scale=0.62]{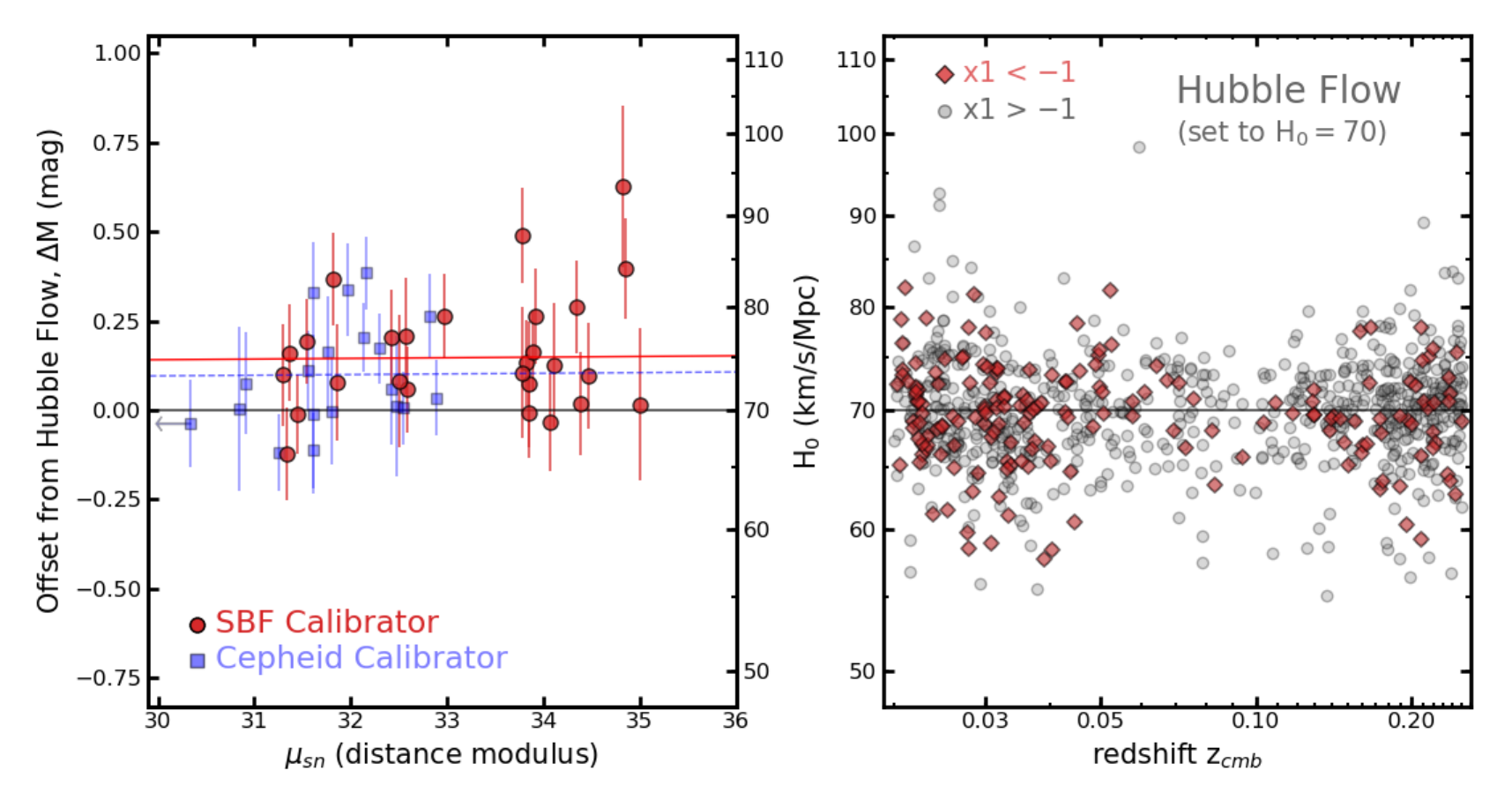}
    \caption{\hnot\ estimated from SBF and Cepheid calibrators using the Pantheon+ light curves anchored by the Hubble flow.
    {\bf left panel:} Each SBF-calibrated event (red points) provides an estimate of \hnot\ based on the offset between the SBF distance and the SALT2 distance. The blue points are the same but for the Cepheid-calibrated \snia\ from \citet{riess16}. Both the SBF- and Cepheid-calibrated events suggest a value for \hnot\ greater than 70~\kmsmpc. The average of the \hnot\ estimates from SBF-calibrated \snia\ is shown as the red solid line. SN~2011fe has been shifted in distance to make the plot more readable.
    {\bf right panel:} The Hubble flow supernovae for $0.02< z_{cmb} < 0.25$, displaying the scatter about their mean luminosity. The luminosity has been set to provide an $H_{0}=70$~\kmsmpc. The light red points indicate supernovae with light curve stretch values of $x1 < -1.0$, similar to the \snia\ from SBF hosts.   }
    \label{hubble2}
\end{figure*}

\subsubsection{The Hubble Flow Distances}

\citet{riess09} outlined the steps needed to calibrate \snia\ occurring in Cepheid-calibrated host galaxies and apply that to the Hubble flow.  Cosmological effects become important in the Hubble flow, so the equation relating the local \hubble\ parameter with redshift includes deceleration/acceleration correction terms in the form of expansions in redshift. We modify Equation~14 from \citet{riess09} to directly use the distance moduli output by SALT2, $\mu_{s2}$. These SALT2 distances have been calculated based on Equation~1 with the fiducial \snia\ absolute magnitude, $M_B$, assigned an arbitrary value. Thus, the $\mu_{s2}$ values can be shifted by a constant, $\Delta M$, that is equivalent to varying $M_B$. That is, the SBF (or Cepheid) calibrators are used to estimate $\Delta M$ and from it, $H_0$.  We define a parameter, $\log H_{s2}$, based on the Hubble flow SALT2 distance moduli: 

%\begin{equation}
\begin{multline} \label{equ_hubble1}
\log{H_{s2}} = \log{c z}\bigg\{ 1 + \frac{1}{2} [1 - q_{0}] z - \frac{1}{6} [1 - q_{0} - 3q_{0}^{2} \\ + j_{0}] z^{2}\bigg\} - 0.2 \mu_{s2} + 5 \;,
\end{multline}
%\end{equation}
where $z$ is the redshift in the CMB frame, $q_0$, is the deceleration parameter, and $j_0$ is the jerk parameter. We set $q_0=-0.54\pm0.06$ with $j_0$ fixed at $1.0$ \citep{riess21}.
%We set $q_0=-0.56\pm0.06$ and $j_0=0.65\pm0.4$ from Brout et al. (2022; in prep).
Each supernova in the Hubble flow provides a $\log H_{s2}$ value, and a weighted average is calculated. The resulting average $\log H_{s2}$ for each sub-sample is given at the bottom of Table~\ref{samples}.

The difference between each calibrator distance modulus, $\mu_{calib}$, and its corresponding supernova SALT2 distance modulus, $\mu_{s2}$, provides the magnitude offset, $\Delta M=\mu_{s2}-\mu_{calib}\,$, from the arbitrary $M_B$ assumed during the SALT2 fitting. Working directly with distance moduli, each \snia\ in a calibrated host provides an estimate of \hnot\ through,

\begin{equation} \label{equ3}
\log{H_{0}^{i}} = \log{H_{s2}} + 0.2(\mu_{s2}^{i} - \mu_{calib}^{i}) \;,
\end{equation}
where the superscript, $i$, indicates the individual calibrators. The estimated $H_0$ and statistical uncertainties from the SBF calibrators, the Cepheid calibrators, and the various sub-samples are given at the top of Table~\ref{samples}.

The parameter $\log H_{s2}$ depends only upon the SALT2 supernova distances in the Hubble flow and its variation between supernova sub-samples provides a test of systematic errors in the Pantheon+ light curve fits and bias corrections. As seen in Table~\ref{samples}, the parameter does not vary significantly for the different \snia\ Hubble flow subsets. This suggests that any systematic differences between the fast ($x1<-1$) and slow ($x1>-1$) subsets have been mitigated through the bias corrections from the SNANA simulations and applied in the Pantheon+ compilation. Because of the small number of events in the fast sub-sample ($x1<-1$), the uncertainty on $\log H_{s2}$ is significantly larger than for the other samples.

\begin{deluxetable*}{lcccccc}
\centering
%\tablewidth{0.5pt}
\tablecaption{$H_0$ from \snia\ Samples \label{samples}}
\tablehead{
\colhead{Calibration} & \colhead{Hubble Flow$^a$} & \colhead{Number of} & \colhead{Reduced} & \colhead{$\sigma$} & \colhead{$\overline{\mu_{s2}-\mu_{calib}}$} & \colhead{$H_0$$^b$}   \\
\colhead{Sample} & \colhead{Sample} & \colhead{\snia} & \colhead{$\chi_\nu^2$} & \colhead{(mag)} & \colhead{(mag)} & \colhead{(\kmsmpc )} \\
\hline %\vspace{-0.3cm}
SBF (Full)    & {Full}     & 27 & 1.25 & 0.153 &  0.154$\pm 0.027$ & 74.61$\pm 0.93$ \\  %jan4.  0.92+0.14
SBF ($x1<-1$)   & {Fast}      & 24 & 1.37 & 0.160 &  0.162$\pm 0.028$   & 74.80$\pm 1.03 $ \\ %jan4. 0.98+0.31
Cepheids (Full) & {Full}     & 19 & 1.61 & 0.153 & 0.109$\pm 0.028$   & 73.07$\pm 0.96$ \\ %jan11. 0.95+0.14
Cepheids ($x1>-1$) & {Slow}     & 17 & 1.75 & 0.164 & 0.100$\pm 0.031$  & 72.75$\pm 1.04 $ \\ %jan4. 1.03+0.16
\hline
\hline
\\
\\
\hline
\colhead{Hubble Flow$^c$} & \colhead{Number of} & \colhead{Reduced} & \colhead{$\sigma$} & \colhead{log$H_{s2}$$^d$} & \colhead{ }  & \colhead{ } \\
\colhead{Sample} & \colhead{\snia} & \colhead{$\chi_\nu^2$} & \colhead{(mag)} & \colhead{ } & \colhead{ } & \colhead{ } }
\startdata
Full & 816 & 1.12 & 0.135 & 1.8420$\pm0.0009$ & & \\ %jan4
Fast: $x1<-1.0$ & 175 & 1.01 & 0.144 & 1.8415$\pm0.0020$ & & \\ %jan4
Slow: $x1>-1.0$ & 641 & 1.16 & 0.152 & 1.8421$\pm0.0010$ & & % jan4
\enddata
\tablenotetext{\tiny a}{Indicates the Hubble Flow sample used to estimate $\log H_{s2}$}
\tablenotetext{\tiny b}{Uncertainties are statistical only and include the error from the Hubble Flow}
\tablenotetext{\tiny c}{The Hubble flow redshift range: 0.02 $< z < 0.25$}
\tablenotetext{\tiny d}{The $\log H_{s2}$ parameter is defined in Equations~\ref{equ_hubble1} and \ref{equ3}.}
\end{deluxetable*}

\section{Results}

\subsection{Full Sample}

\hnot\ estimates from the SBF and Cepheid calibrators are displayed in the left panel of Figure~\ref{hubble2}. The SBF calibrator distances average 0.15~mag fainter than the SALT2 \snia\ distance moduli when the $M_B$ parameter is set to yield $70.0$~\kmsmpc\ in the Hubble flow (right panel of Figure~\ref{hubble2}). More precisely, the SBF-calibrators applied to the full Hubble flow supernova sample provide an estimate of $H_0=74.61$~\kmsmpc\ with a statistical uncertainty of $\pm 0.93$~\kmsmpc\ as shown in Table~\ref{samples}. The Cepheid calibrators applied to the Hubble flow supernovae yield $H_0=73.07\pm 0.96$~\kmsmpc , and this is consistent with the measurement by \citet{riess21}. We note that Figure~\ref{hubble2} shows that the SBF-calibrators extend two magnitudes further in distance modulus than the Cepheid-calibrated \snia, thus, the SBF calibrators reach the near edge of the Hubble flow as defined here.

%The $\chi^2$ parameter for all the data sets is too large by a factor of about 1.5, suggesting that the calibrator distance uncertainties are underestimated. 

\subsection{Fast vs. Slow Sub-samples}

Using only fast-declining \snia\ results in an insignificant shift in the estimated \hnot\ value when compared with the Full sample. Table~\ref{samples} shows that the 175 Hubble flow events with $x1<-1.0$ and calibrated with IR~SBF distances shift the \hnot\ estimate by less than 0.2~\kmsmpc\ . Calibrating the slow-fading \snia\ with Cepheid distances also has a minimal impact on the results.

\subsection{SBF-like Sub-sample}

In the previous analyses, we applied calibrations using supernova distances conditioned using bias corrections derived from SNANA simulations \citep{pantheon+2}. This approach works very well as demonstrated by the consistent results when sub-samples of the Hubble flow supernovae are analyzed. But there remains concern that the unique character of \snia\ found in SBF-calibrated galaxy (``SBF-like'' supernovae defined in Section~\ref{subsets}) may not be fully accounted for in the simulations. For the 162 Hubble flow \snia\ in the SBF-like sub-sample, we refit the SALT2 $\alpha$ and $\beta$ parameters so as to minimize their Hubble residuals. In this sub-sample, the typical stretch parameter value is $x1 <-1$, so we recenter the stretch distribution using a modified Tripp relation:
\begin{equation}
\mu\; =\; m_B+\alpha\;(x1+1.5)-\beta\;c - M_B\;\; ,
\end{equation}
which avoids creating a large correlation between $\alpha$ and $M_B$.
We then estimate \hnot\ by applying the new SALT2 coefficients to IR~SBF-calibrated \snia\ light curves and averaging the results from Equation~\ref{equ3}.

\begin{figure*}
    \centering
    \includegraphics[scale=0.6]{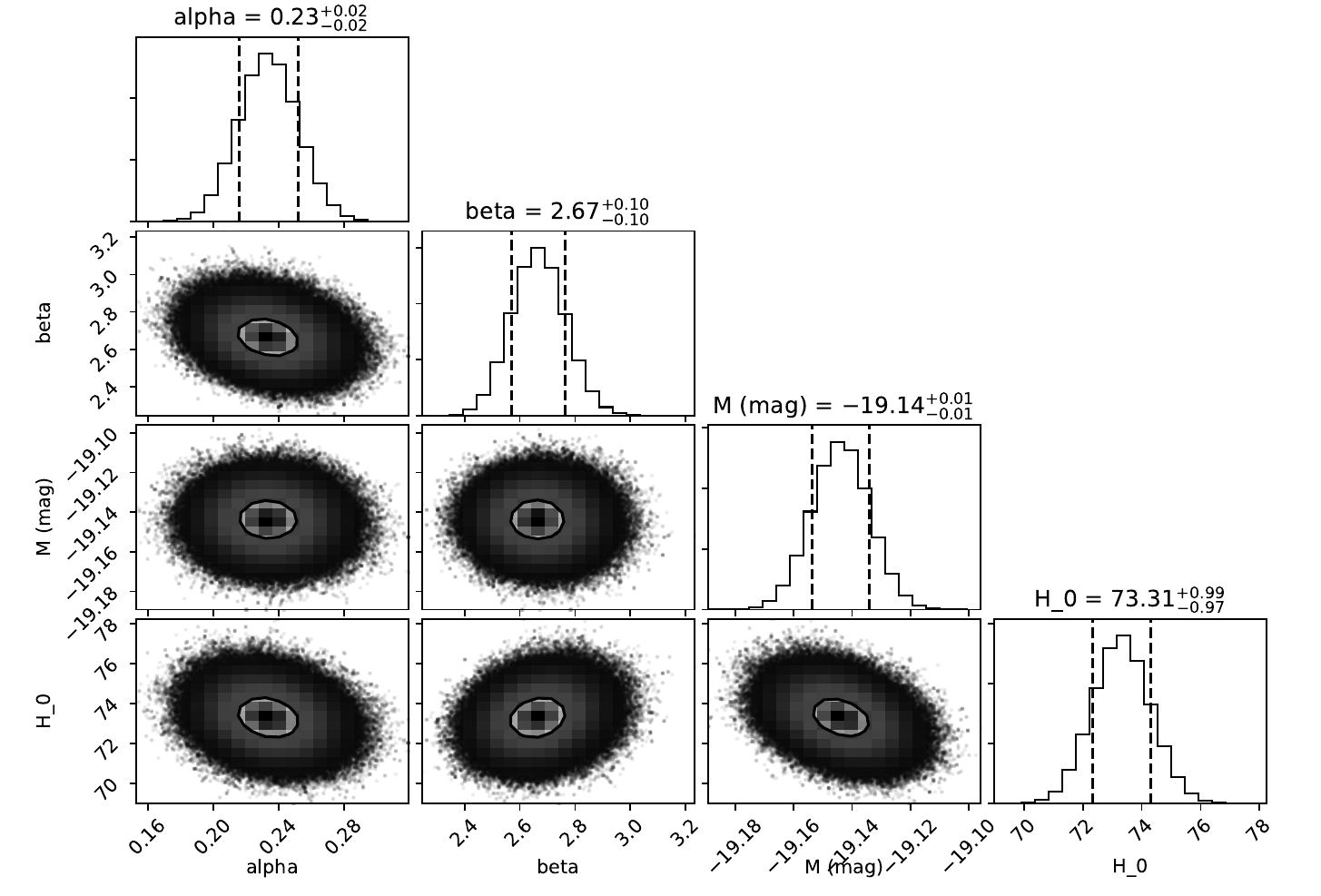}
    \caption{The MCMC results for refitting the SALT2 parameters using Hubble flow \snia\ with fast-declining light curves in massive hosts (SBF-like events). The revised SALT2 parameters are then applied to the SBF-calibrated \snia\ and \hnot\ is estimated from the IR~SBF distances. The resulting \hnot\ estimate is consistent with that from Cepheid-calibrated \snia. }
    \label{corner}
\end{figure*}

We ran a Markov-Chain Monte Carlo (MCMC) calculation on the SBF-like \hubble\ flow supernovae to determine the SALT2 parameters that minimize the Hubble residuals. The MCMC code then applies these parameters to the 24 \snia\ in the SBF calibration sample that meet the stretch and host mass criteria. The SALT2 parameters and the resulting \hnot\ are shown in Figure~\ref{corner}. As expected, the luminosity slope ($\alpha = 0.23\pm 0.02$) is found to be very steep when compared with the slope for the slow-decliners. The color correction parameter ($\beta=2.7\pm0.1$) is similar to that found for the Full sample. 

The fiducial absolute magnitude after refitting, $M_B=-19.14$, is fainter than that shown in Figure~\ref{phillips} because here we reference to a stretch parameter of $x1=-1.5$ instead of $x1=0.0$. Applying these parameters to the Hubble flow supernovae and the \snia\ in the SBF calibrators results in an \hnot\ value that is 1.5~\kmsmpc\ smaller than the results from the Pantheon+ distances, but within the statistical uncertainties of the two methods. The $H_0=73.31\pm 0.99$(stat) estimated from the SBF-like sample with refitted SALT2 coefficients is nearly identical to that found by \citet{blakeslee21} using IR~SBF measurements as their ``top rung.'' Both are in excellent agreement with the \citet{riess21} Cepheid measurement of $H_0$ (see Figure~\ref{summary}).

\subsection{The \hubble\ Constant } \label{sec:hubble}

As shown in the previous sections, the IR~SBF distance measured to \snia\ hosts allows us to calibrate the Hubble flow supernovae and estimate the \hubble\ constant with a zero point established from Cepheid variables. Using the full Hubble flow sample of Pantheon+ supernova distances results in $H_0=74.61$~\kmsmpc\ with a statistical uncertainty of 0.93~\kmsmpc. 

Dividing the \snia\ into subsamples based on light curve and host properties moves the calibration at the 0.03~mag level. Thus, we include a systematic uncertainty of 1.8\%\ on the $H_0$ estimate caused by the differences between \snia\ found in SBF-calibrated hosts and supernovae exploding in star-forming host galaxies. In addition, the zero point of the IR~SBF distances has a systematic uncertainty of 3.1\%\ based on the \citet{blakeslee21} SBF calibration. A summary of statistical and systematic errors contributing to our \hubble\ constant estimate is given in Table~\ref{errors}.

The IR~SBF/\snia\ derived \hubble\ constant is consistent with the Cepheid/\snia\ based on the \citet{riess21} distances. Because the IR~SBF zero point is based on Cepheids \citep{jensen21}, we expect that the 2\%\ shift in $H_0$ (Table~3) may be a result of the extreme differences in \snia\ properties between star-forming hosts and massive, passive hosts. Indeed, the largest offset ($\sim$3\%)  is seen between IR~SBF-calibrated hosts using only fast-fading supernovae, and the Cepheid-calibrated hosts using only slow-fading events.

\section{Discussion}

\snia\ in SBF-calibrated hosts appear quite different from those in Cepheid-calibrated hosts. The majority of light curves in the SBF-calibrated sample are narrow ($x1<-1.0$) and are found in hosts with stellar masses greater than 10$^{10.5}$~M$_\odot$. These differences suggest a possible systematic shift between SBF and Cepheid calibrations, but no significant systematic is found based on the consistency in \hnot\ estimated from various sub-samples of the Pantheon+ \snia .

%**********************
%***********************

The SNANA simulations and resulting bias corrections of the \snia\ light curve parameters appears to have successfully mitigated significant systematic differences between the SBF- and Cepheid-calibrated \snia\ populations. Further, our test of isolating SBF-like \snia\ and refitting SALT2 coefficients also appears to have resulted in a calibration consistent with the calibration of the entire \snia\ sample. The SBF method as a calibrator of the Hubble flow \snia\ does not appear biased relative to other methods. However, the IR~SBF distance scale is calibrated using a Cepheid zero point, so is not completely independent of other methods, and is subject to the same common systematic uncertainties at lower rungs of the distance ladder.

The physical origin of the extreme differences between \snia\ occurring in SBF-calibrated hosts and Cepheid-calibrated hosts remains uncertain. The difference in light curve width between elliptical and spiral host galaxies is well known \citep{hamuy96,gallagher05}, and maybe more evident here because the SBF method selects the most extreme hosts in age and mass. This is contrasted with the Cepheid-calibrated \snia\ that tend to originate in hosts that require some recent star formation and are lower in mass than the passive hosts. There is certainly a formation-to-explosion delay-time difference between the SBF-calibrated and Cepheid-calibrated supernovae. It remains unclear if all \snia\ events arise from a single progenitor whose explosion properties evolve with age or result from multiple progenitor mechanisms that overlap in their observable properties.

Excluding the fast decliners, Figure~\ref{phillips_mass} clearly shows that the average peak luminosity from the SALT2 analysis of light curves is 0.1~mag brighter for \snia\ in high-mass hosts, which is the origin of the ``mass-step'' correction. \citet{brout21} have argued that this is due to changes in dust properties with increasing galaxy mass. %However, the existence of a mass-step correction for the fast decliners is not as obvious. The massive, passive hosts that dominate the SBF-like sub-sample are generally not expected to contain as much dust as spiral galaxies.
A recent study by \citet{chen22} suggests that the relation between color and luminosity for \snia\ found in massive passive hosts is flatter than seen in star-forming galaxies.

For the fast-declining \snia\ shown in Figure~\ref{phillips_mass}, there is a clear increase in the $\alpha$ parameter moving from low to high mass hosts. In fact, the absolute peak magnitude of supernovae with stretch $x1 < -1$ are best fit with an extremely steep $\alpha = 0.31\pm 0.05$. We are likely seeing an increasing fraction of low-luminosity supernovae in the most massive hosts. While we have culled events that are spectroscopically identified as SN1991bg-like, there may be a transition between fast-declining ``narrow normals'' and true SN1991bg-like events that results in the steepening slope seen in the Phillips relation. This transition could be identified with the \snia\ in the lower-right of Figure~\ref{color_stretch} where there is a cluster of reddish/fast-decliners in massive hosts. Our lowest luminosity \snia\ in the SBF calibration sample, SN2016ajf, is likely a member of this group. The fact that SN2016ajf passed all the SNANA bias cuts suggests that these transition events may still useful as distance indicators and calibrators.

\begin{deluxetable}{lccc}
\centering
\tablewidth{0.5pt}
\tablecaption{Uncertainties on $H_0$ from IR~SBF \label{errors}}
\tablehead{
\colhead{Source} & \colhead{Type} & \colhead{$\sigma$ (mag)} & \colhead{$H_0$} 
}
\startdata
IR SBF/\snia\ hosts & stat  & 0.027 &  1.2\%  \\
Hubble flow & stat  & 0.005 &  0.2\% \\
\snia\ samples & syst  & 0.038 & 1.8\%  \\
Zero point$^a$ & syst   &  0.066 & 3.1\%  \\   %3.1% from Blakeslee
\hline
Total  &  stat   &  0.027 &  1.2\%  \\
Total  &  syst   &  0.076 & 3.6\%  \\
\enddata
\tablenotetext{\tiny a}{From \citet{blakeslee21}.}
\end{deluxetable}

\section{Conclusion}

We have used new IR-SBF distances to 25 unique \snia\ host galaxies to calibrate the luminosity of a large set of Hubble flow \snia. We have tested the robustness of the calibration in three ways. First, we used Pantheon+ bias-corrected distances to estimate the \hubble\ constant from SBF and Cepheid calibrators. Second, we divided the Hubble flow sample into sub-samples with properties similar to that of the calibrator samples. Finally, we identified supernovae in the Hubble flow that simultaneously matched the stretch parameters and host galaxy masses of the SBF calibrators. We then refit their light curve correction coefficients to obtain distances that were independent of the Pantheon+ bias corrections. We then searched for systematic shifts in the \hubble\ constants derived from these techniques. A summary of our results is shown in Figure~\ref{summary} and described here:

\begin{itemize}

\item Directly comparing the SALT2 \snia\ distances and IR~SBF distances demonstrates that the estimated uncertainties given in \citet{jensen21} accurately reflect the observed scatter. Distance uncertainties to individual IR~SBF hosts are comparable to uncertainties in the Cepheid calibrators.

%There remains some concern with inconsistencies in individual hosts (e.g. NGC1316) and supernovae (e.g. SN2007on and SN2011iv).

\item The light curve widths for the SBF-calibrated \snia\ are typically extremely narrow when compared to the Cepheid-calibrated \snia, with 24 of 27 of the SBF-calibrated \snia\ having a stretch parameter of $x1<-1.0$ while only 2 in 19 of the Cepheid-calibrated \snia\ are seen to fade so quickly . The distribution of \snia\ colors at peak is similar between SBF- and Cepheid-calibrated \snia\ when SN1991bg-like events are excluded. 

%**********************
\begin{figure}
    \centering
    \includegraphics[scale=0.53]{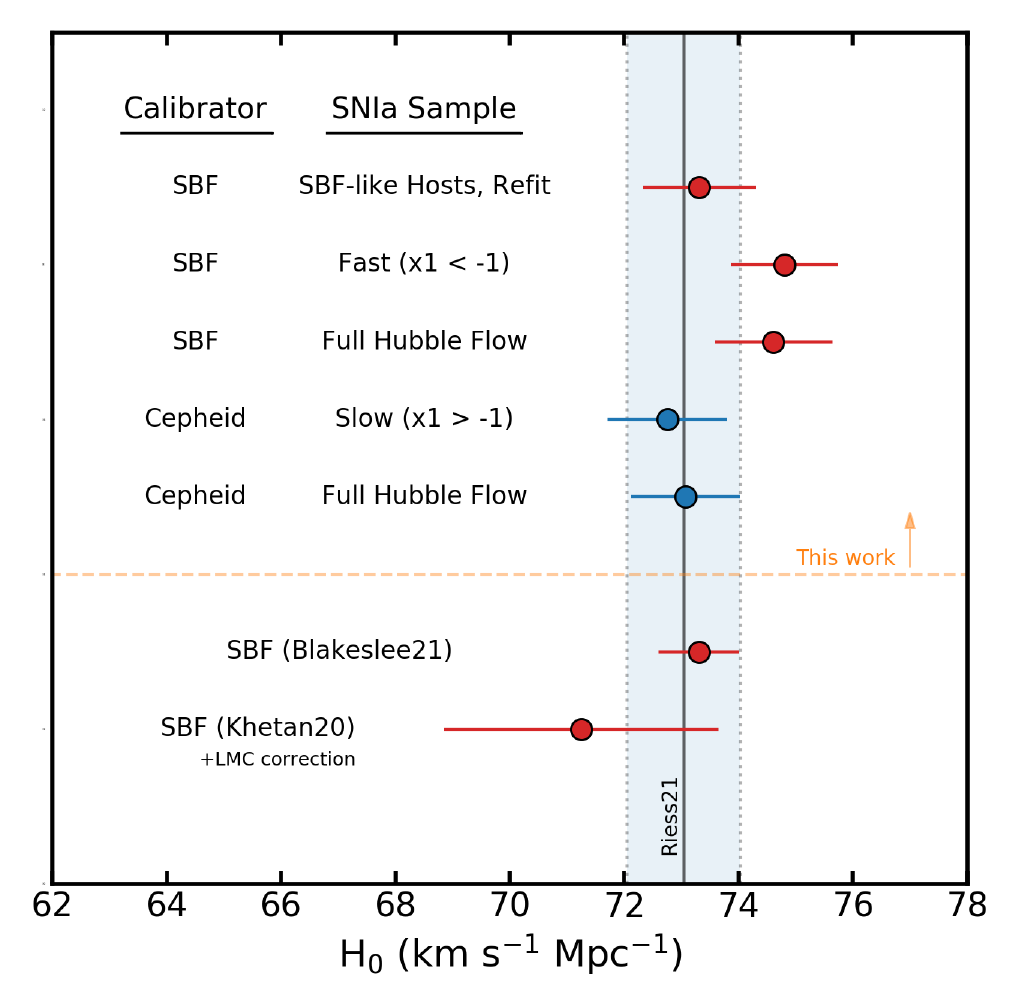}
    \caption{The results of this work compared with \hubble\ constant estimates using SBF from \citet{khetan21} and \citet{blakeslee21}. Errorbars on points indicate statistical uncertainties only. The light blue band is the \citet{riess21} estimate using Cepheids (indicating combined statistical and systematic uncertainties).  The top-most point shows the results from fitting supernova luminosity corrections trained on fast-declining \snia\ occurring in massive host galaxies (SBF-like events). Our other results use the Pantheon+ bias-corrected SALT2 \snia\ distances. }
    \label{summary}
\end{figure}
%***********************

\item The majority of SBF-calibrator galaxies possess larger stellar masses than the star-forming spirals calibrated by Cepheids. All but one of the SBF-calibrator galaxies has a stellar mass above the typical location of the ``mass-step'' at 10$^{10}$~M$_\odot$, while the median galaxy mass of Cepheid-calibrators is just below 10$^{10}$~M$_\odot$. 

\item The relation between \snia\ peak brightness and light curve width appears to steepen with decreasing stretch (SALT2 $x1$), suggesting that the linear Tripp relation should be modified when applied to fast declining events. Other fitters \citep[e.g.][]{phillips99, burns18} have seen such a non-linear relationship between luminosity and light curve width. 

\item We find the difference between the IR-SBF calibration and Cepheid calibration of the Hubble flow supernovae to be 0.045$\pm 0.038$~mag or a 2.0$\pm 1.7$\%\ difference in H$_0$.

\item When the sample was restricted to \snia\ with narrow light curves ($x1 < -1.0$, the ``fast'' sub-sample),  the estimated \hubble\ constant changed by only 0.19~\kmsmpc\ compared with the estimate that used the full range of stretch. Thus, there is no significant shift between the full Hubble flow and the sample containing only fast fading supernovae. We conclude that bias corrections derived from SNANA simulations are effective at reducing systematic distance residuals across the full range of light curve decline rates.

\item Refitting SALT2 parameters using only \snia\ in the ``SBF-like'' sub-sample (selected by both host mass and stretch parameter), we find a steep luminosity slope $\alpha =0.23\pm 0.02$ best corrects the \snia\ luminosities. The \hubble\ constant derived from the revised SALT2 parameter fit and calibrated with IR~SBF distances is $H_0=73.3\pm 1.0$~\kmsmpc , or within 0.3\%\ of the H$_0$ derived from the Cepheid-calibrated \snia\ sample. 

\item Using IR~SBF distances to calibrate the Pantheon+ \snia\ and then applying this calibration to Pantheon+ \snia\ in the  Hubble flow, we estimate \hnot$= 74.6\pm 0.9$(stat)$\pm 2.7$(syst)~\kmsmpc . The IR~SBF zero point is based on Cepheid distances, so this is not a fully independent test of the Hubble tension.  But, the consistency between the IR SBF and Cepheid calibrators suggests that the top rung of the \snia\ distance ladder is not a major source of the Hubble tension.

\end{itemize}

We find that the IR SBF method for calibrating distances is competitive with the Cepheid calibration of nearby \snia. Despite the divergent properties we identify between \snia\ found in star-forming and massive/passive hosts,  we find no significant systematic difference between the SBF and Cepheid calibrators when using the bias-corrected Pantheon+ distances or refitting the luminosity/stretch relation for fast fading \snia . 

Compared to Cepheid calibrations, the IR SBF method can reach more distant \snia\ hosts with a lower observing time cost per calibrator.  The angular resolution of JWST is likely to boost the reach of IR-SBF distance measurements. Nearby, the JWST calibration of distance indicators found in old populations, such as the TRGB, may lead to a robust distance ladder out to the Hubble flow that is independent of Cepheids and further test the origin of the Hubble tension.

\begin{acknowledgements}

We thank Rebecca Chen for comparing these results to the Khetan et al. distances.

\end{acknowledgements}

%\appendix
%\section{ Luminosity-Stretch Correction for \snia\ Subsets}

\bibliography{bib.bib}

\begin{thebibliography}{}
\expandafter\ifx\csname natexlab\endcsname\relax\def\natexlab#1{#1}\fi
\providecommand{\url}[1]{\href{#1}{#1}}
\providecommand{\dodoi}[1]{doi:~\href{http://doi.org/#1}{\nolinkurl{#1}}}
\providecommand{\doeprint}[1]{\href{http://ascl.net/#1}{\nolinkurl{http://ascl.net/#1}}}
\providecommand{\doarXiv}[1]{\href{https://arxiv.org/abs/#1}{\nolinkurl{https://arxiv.org/abs/#1}}}

\bibitem[{{Ashall} {et~al.}(2018){Ashall}, {Mazzali}, {Stritzinger},
  {Hoeflich}, {Burns}, {Gall}, {Hsiao}, {Phillips}, {Morrell}, \&
  {Foley}}]{ashall18}
{Ashall}, C., {Mazzali}, P.~A., {Stritzinger}, M.~D., {et~al.} 2018, \mnras,
  477, 153, \dodoi{10.1093/mnras/sty632}

\bibitem[{{Ashall} {et~al.}(2020){Ashall}, {Lu}, {Burns}, {Hsiao},
  {Stritzinger}, {Suntzeff}, {Phillips}, {Baron}, {Contreras}, {Davis},
  {Galbany}, {Hoeflich}, {Holmbo}, {Morrell}, {Karamehmetoglu}, {Krisciunas},
  {Kumar}, {Shahbandeh}, \& {Uddin}}]{ashall20}
{Ashall}, C., {Lu}, J., {Burns}, C., {et~al.} 2020, \apjl, 895, L3,
  \dodoi{10.3847/2041-8213/ab8e37}

\bibitem[{{Bernal} {et~al.}(2016){Bernal}, {Verde}, \& {Riess}}]{bernal16}
{Bernal}, J.~L., {Verde}, L., \& {Riess}, A.~G. 2016, \jcap, 2016, 019,
  \dodoi{10.1088/1475-7516/2016/10/019}

\bibitem[{{Betoule} {et~al.}(2014){Betoule}, {Kessler}, {Guy}, {Mosher},
  {Hardin}, {Biswas}, {Astier}, {El-Hage}, {Konig}, {Kuhlmann}, {Marriner},
  {Pain}, {Regnault}, {Balland}, {Bassett}, {Brown}, {Campbell}, {Carlberg},
  {Cellier-Holzem}, {Cinabro}, {Conley}, {D'Andrea}, {DePoy}, {Doi}, {Ellis},
  {Fabbro}, {Filippenko}, {Foley}, {Frieman}, {Fouchez}, {Galbany}, {Goobar},
  {Gupta}, {Hill}, {Hlozek}, {Hogan}, {Hook}, {Howell}, {Jha}, {Le Guillou},
  {Leloudas}, {Lidman}, {Marshall}, {M{\"o}ller}, {Mour{\~a}o}, {Neveu},
  {Nichol}, {Olmstead}, {Palanque-Delabrouille}, {Perlmutter}, {Prieto},
  {Pritchet}, {Richmond}, {Riess}, {Ruhlmann-Kleider}, {Sako}, {Schahmaneche},
  {Schneider}, {Smith}, {Sollerman}, {Sullivan}, {Walton}, \&
  {Wheeler}}]{betoule14}
{Betoule}, M., {Kessler}, R., {Guy}, J., {et~al.} 2014, \aap, 568, A22,
  \dodoi{10.1051/0004-6361/201423413}

\bibitem[{{Blakeslee} {et~al.}(2021){Blakeslee}, {Jensen}, {Ma}, {Milne}, \&
  {Greene}}]{blakeslee21}
{Blakeslee}, J.~P., {Jensen}, J.~B., {Ma}, C.-P., {Milne}, P.~A., \& {Greene},
  J.~E. 2021, \apj, 911, 65, \dodoi{10.3847/1538-4357/abe86a}

\bibitem[{{Blakeslee} {et~al.}(2009){Blakeslee}, {Jord{\'a}n}, {Mei},
  {C{\^o}t{\'e}}, {Ferrarese}, {Infante}, {Peng}, {Tonry}, \&
  {West}}]{blakeslee09}
{Blakeslee}, J.~P., {Jord{\'a}n}, A., {Mei}, S., {et~al.} 2009, \apj, 694, 556,
  \dodoi{10.1088/0004-637X/694/1/556}

\bibitem[{{Blakeslee} {et~al.}(2010){Blakeslee}, {Cantiello}, {Mei},
  {C{\^o}t{\'e}}, {Barber DeGraaff}, {Ferrarese}, {Jord{\'a}n}, {Peng},
  {Tonry}, \& {Worthey}}]{blakeslee10}
{Blakeslee}, J.~P., {Cantiello}, M., {Mei}, S., {et~al.} 2010, \apj, 724, 657,
  \dodoi{10.1088/0004-637X/724/1/657}

\bibitem[{{Brout} \& {Scolnic}(2021)}]{brout21}
{Brout}, D., \& {Scolnic}, D. 2021, \apj, 909, 26,
  \dodoi{10.3847/1538-4357/abd69b}

\bibitem[{{Brout} {et~al.}(2022){Brout}, {Taylor}, {Scolnic}, {Wood}, {Rose},
  {Vincenzi}, {Dwomoh}, {Lidman}, {Riess}, {Ali}, {Qu}, \& {Dai}}]{pantheon+2}
{Brout}, D., {Taylor}, G., {Scolnic}, D., {et~al.} 2022, \apj, 938, 111,
  \dodoi{10.3847/1538-4357/ac8bcc}

\bibitem[{{Brown} {et~al.}(2018){Brown}, {Perry}, {Beeny}, {Milne}, \&
  {Wang}}]{Swift}
{Brown}, P.~J., {Perry}, J.~M., {Beeny}, B.~A., {Milne}, P.~A., \& {Wang}, X.
  2018, \apj, 867, 56, \dodoi{10.3847/1538-4357/aae1ad}

\bibitem[{{Burns} {et~al.}(2011){Burns}, {Stritzinger}, {Phillips}, {Kattner},
  {Persson}, {Madore}, {Freedman}, {Boldt}, {Campillay}, {Contreras},
  {Folatelli}, {Gonzalez}, {Krzeminski}, {Morrell}, {Salgado}, \&
  {Suntzeff}}]{burns11}
{Burns}, C.~R., {Stritzinger}, M., {Phillips}, M.~M., {et~al.} 2011, \aj, 141,
  19, \dodoi{10.1088/0004-6256/141/1/19}

\bibitem[{{Burns} {et~al.}(2018){Burns}, {Parent}, {Phillips}, {Stritzinger},
  {Krisciunas}, {Suntzeff}, {Hsiao}, {Contreras}, {Anais}, {Boldt}, {Busta},
  {Campillay}, {Castell{\'o}n}, {Folatelli}, {Freedman}, {Gonz{\'a}lez},
  {Hamuy}, {Heoflich}, {Krzeminski}, {Madore}, {Morrell}, {Persson}, {Roth},
  {Salgado}, {Ser{\'o}n}, \& {Torres}}]{burns18}
{Burns}, C.~R., {Parent}, E., {Phillips}, M.~M., {et~al.} 2018, \apj, 869, 56,
  \dodoi{10.3847/1538-4357/aae51c}

\bibitem[{{Cantiello} {et~al.}(2013){Cantiello}, {Grado}, {Blakeslee},
  {Raimondo}, {Di Rico}, {Limatola}, {Brocato}, {Della Valle}, \&
  {Gilmozzi}}]{cantiello13}
{Cantiello}, M., {Grado}, A., {Blakeslee}, J.~P., {et~al.} 2013, \aap, 552,
  A106, \dodoi{10.1051/0004-6361/201220756}

\bibitem[{{Cantiello} {et~al.}(2018){Cantiello}, {Blakeslee}, {Ferrarese},
  {C{\^o}t{\'e}}, {Roediger}, {Raimondo}, {Peng}, {Gwyn}, {Durrell}, \&
  {Cuillandre}}]{cantiello18}
{Cantiello}, M., {Blakeslee}, J.~P., {Ferrarese}, L., {et~al.} 2018, \apj, 856,
  126, \dodoi{10.3847/1538-4357/aab043}

\bibitem[{{Chen} {et~al.}(2022){Chen}, {Scolnic}, {Rozo}, {Rykoff}, {Popovic},
  {Kessler}, {Vincenzi}, {Davis}, {Armstrong}, {Brout}, {Galbany}, {Kelsey},
  {Lidman}, {M{\"o}ller}, {Rose}, {Sako}, {Sullivan}, {Taylor}, {Wiseman},
  {Asorey}, {Carr}, {Conselice}, {Kuehn}, {Lewis}, {Macaulay},
  {Rodriguez-Monroy}, {Tucker}, {Abbott}, {Aguena}, {Allam},
  {Andrade-Oliveira}, {Annis}, {Bacon}, {Bertin}, {Bocquet}, {Brooks}, {Burke},
  {Carnero Rosell}, {Carrasco Kind}, {Carretero}, {Cawthon}, {Costanzi}, {da
  Costa}, {Pereira}, {Desai}, {Diehl}, {Doel}, {Everett}, {Ferrero},
  {Flaugher}, {Friedel}, {Frieman}, {Garc{\'\i}a-Bellido}, {Gatti},
  {Gaztanaga}, {Gruen}, {Hinton}, {Hollowood}, {Honscheid}, {James}, {Lahav},
  {Lima}, {March}, {Menanteau}, {Miquel}, {Morgan}, {Palmese},
  {Paz-Chinch{\'o}n}, {Pieres}, {Plazas Malag{\'o}n}, {Prat}, {Romer},
  {Roodman}, {Sanchez}, {Schubnell}, {Serrano}, {Sevilla-Noarbe}, {Smith},
  {Soares-Santos}, {Suchyta}, {Tarle}, {Thomas}, {To}, {Tucker}, \&
  {Varga}}]{chen22}
{Chen}, R., {Scolnic}, D., {Rozo}, E., {et~al.} 2022, \apj, 938, 62,
  \dodoi{10.3847/1538-4357/ac8b82}

\bibitem[{{Childress} {et~al.}(2014){Childress}, {Wolf}, \&
  {Zahid}}]{childress14}
{Childress}, M.~J., {Wolf}, C., \& {Zahid}, H.~J. 2014, \mnras, 445, 1898,
  \dodoi{10.1093/mnras/stu1892}

\bibitem[{{Efstathiou}(2021)}]{efstathiou21}
{Efstathiou}, G. 2021, \mnras, 505, 3866, \dodoi{10.1093/mnras/stab1588}

\bibitem[{{Foley} {et~al.}(2018){Foley}, {Scolnic}, {Rest}, {Jha}, {Pan},
  {Riess}, {Challis}, {Chambers}, {Coulter}, {Dettman}, {Foley}, {Fox},
  {Huber}, {Jones}, {Kilpatrick}, {Kirshner}, {Schultz}, {Siebert},
  {Flewelling}, {Gibson}, {Magnier}, {Miller}, {Primak}, {Smartt}, {Smith},
  {Wainscoat}, {Waters}, \& {Willman}}]{FOUND}
{Foley}, R.~J., {Scolnic}, D., {Rest}, A., {et~al.} 2018, \mnras, 475, 193,
  \dodoi{10.1093/mnras/stx3136}

\bibitem[{{Freedman} {et~al.}(2001){Freedman}, {Madore}, {Gibson}, {Ferrarese},
  {Kelson}, {Sakai}, {Mould}, {Kennicutt}, {Ford}, {Graham}, {Huchra},
  {Hughes}, {Illingworth}, {Macri}, \& {Stetson}}]{freedman01}
{Freedman}, W.~L., {Madore}, B.~F., {Gibson}, B.~K., {et~al.} 2001, \apj, 553,
  47, \dodoi{10.1086/320638}

\bibitem[{{Freedman} {et~al.}(2019){Freedman}, {Madore}, {Hatt}, {Hoyt},
  {Jang}, {Beaton}, {Burns}, {Lee}, {Monson}, {Neeley}, {Phillips}, {Rich}, \&
  {Seibert}}]{freedman19}
{Freedman}, W.~L., {Madore}, B.~F., {Hatt}, D., {et~al.} 2019, \apj, 882, 34,
  \dodoi{10.3847/1538-4357/ab2f73}

\bibitem[{{Gall} {et~al.}(2018){Gall}, {Stritzinger}, {Ashall}, {Baron},
  {Burns}, {Hoeflich}, {Hsiao}, {Mazzali}, {Phillips}, {Filippenko},
  {Anderson}, {Benetti}, {Brown}, {Campillay}, {Challis}, {Contreras}, {Elias
  de la Rosa}, {Folatelli}, {Foley}, {Fraser}, {Holmbo}, {Marion}, {Morrell},
  {Pan}, {Pignata}, {Suntzeff}, {Taddia}, {Torres Robledo}, \&
  {Valenti}}]{gall18}
{Gall}, C., {Stritzinger}, M.~D., {Ashall}, C., {et~al.} 2018, \aap, 611, A58,
  \dodoi{10.1051/0004-6361/201730886}

\bibitem[{{Gallagher} {et~al.}(2005){Gallagher}, {Garnavich}, {Berlind},
  {Challis}, {Jha}, \& {Kirshner}}]{gallagher05}
{Gallagher}, J.~S., {Garnavich}, P.~M., {Berlind}, P., {et~al.} 2005, \apj,
  634, 210, \dodoi{10.1086/491664}

\bibitem[{{Ganeshalingam} {et~al.}(2010){Ganeshalingam}, {Li}, {Filippenko},
  {Anderson}, {Foster}, {Gates}, {Griffith}, {Grigsby}, {Joubert}, {Leja},
  {Lowe}, {Macomber}, {Pritchard}, {Thrasher}, \& {Winslow}}]{KAITM}
{Ganeshalingam}, M., {Li}, W., {Filippenko}, A.~V., {et~al.} 2010, \apjs, 190,
  418, \dodoi{10.1088/0067-0049/190/2/418}

\bibitem[{{Garnavich} {et~al.}(1998){Garnavich}, {Jha}, {Challis},
  {Clocchiatti}, {Diercks}, {Filippenko}, {Gilliland}, {Hogan}, {Kirshner},
  {Leibundgut}, {Phillips}, {Reiss}, {Riess}, {Schmidt}, {Schommer}, {Smith},
  {Spyromilio}, {Stubbs}, {Suntzeff}, {Tonry}, \& {Carroll}}]{garnavich98}
{Garnavich}, P.~M., {Jha}, S., {Challis}, P., {et~al.} 1998, \apj, 509, 74,
  \dodoi{10.1086/306495}

\bibitem[{{Garnavich} {et~al.}(2004){Garnavich}, {Bonanos}, {Krisciunas},
  {Jha}, {Kirshner}, {Schlegel}, {Challis}, {Macri}, {Hatano}, {Branch},
  {Bothun}, \& {Freedman}}]{garnavich04}
{Garnavich}, P.~M., {Bonanos}, A.~Z., {Krisciunas}, K., {et~al.} 2004, \apj,
  613, 1120, \dodoi{10.1086/422986}

\bibitem[{{Guy} {et~al.}(2005){Guy}, {Astier}, {Nobili}, {Regnault}, \&
  {Pain}}]{guy05}
{Guy}, J., {Astier}, P., {Nobili}, S., {Regnault}, N., \& {Pain}, R. 2005,
  \aap, 443, 781, \dodoi{10.1051/0004-6361:20053025}

\bibitem[{{Guy} {et~al.}(2007){Guy}, {Astier}, {Baumont}, {Hardin}, {Pain},
  {Regnault}, {Basa}, {Carlberg}, {Conley}, {Fabbro}, {Fouchez}, {Hook},
  {Howell}, {Perrett}, {Pritchet}, {Rich}, {Sullivan}, {Antilogus}, {Aubourg},
  {Bazin}, {Bronder}, {Filiol}, {Palanque-Delabrouille}, {Ripoche}, \&
  {Ruhlmann-Kleider}}]{guy07}
{Guy}, J., {Astier}, P., {Baumont}, S., {et~al.} 2007, \aap, 466, 11,
  \dodoi{10.1051/0004-6361:20066930}

\bibitem[{{Hamuy} {et~al.}(2021){Hamuy}, {Cartier}, {Contreras}, \&
  {Suntzeff}}]{hamuy21}
{Hamuy}, M., {Cartier}, R., {Contreras}, C., \& {Suntzeff}, N.~B. 2021, \mnras,
  500, 1095, \dodoi{10.1093/mnras/staa3350}

\bibitem[{{Hamuy} {et~al.}(1995){Hamuy}, {Phillips}, {Maza}, {Suntzeff},
  {Schommer}, \& {Aviles}}]{hamuy95}
{Hamuy}, M., {Phillips}, M.~M., {Maza}, J., {et~al.} 1995, \aj, 109, 1,
  \dodoi{10.1086/117251}

\bibitem[{{Hamuy} {et~al.}(1996){Hamuy}, {Phillips}, {Suntzeff}, {Schommer},
  {Maza}, \& {Aviles}}]{hamuy96}
{Hamuy}, M., {Phillips}, M.~M., {Suntzeff}, N.~B., {et~al.} 1996, \aj, 112,
  2391, \dodoi{10.1086/118190}

\bibitem[{{Hicken} {et~al.}(2009){Hicken}, {Challis}, {Jha}, {Kirshner},
  {Matheson}, {Modjaz}, {Rest}, {Wood-Vasey}, {Bakos}, {Barton}, {Berlind},
  {Bragg}, {Brice{\~n}o}, {Brown}, {Caldwell}, {Calkins}, {Cho}, {Ciupik},
  {Contreras}, {Dendy}, {Dosaj}, {Durham}, {Eriksen}, {Esquerdo}, {Everett},
  {Falco}, {Fernandez}, {Gaba}, {Garnavich}, {Graves}, {Green}, {Groner},
  {Hergenrother}, {Holman}, {Hradecky}, {Huchra}, {Hutchison}, {Jerius},
  {Jordan}, {Kilgard}, {Krauss}, {Luhman}, {Macri}, {Marrone}, {McDowell},
  {McIntosh}, {McNamara}, {Megeath}, {Mochejska}, {Munoz}, {Muzerolle},
  {Naranjo}, {Narayan}, {Pahre}, {Peters}, {Peterson}, {Rines}, {Ripman},
  {Roussanova}, {Schild}, {Sicilia-Aguilar}, {Sokoloski}, {Smalley}, {Smith},
  {Spahr}, {Stanek}, {Barmby}, {Blondin}, {Stubbs}, {Szentgyorgyi}, {Torres},
  {Vaz}, {Vikhlinin}, {Wang}, {Westover}, {Woods}, \& {Zhao}}]{CFA3}
{Hicken}, M., {Challis}, P., {Jha}, S., {et~al.} 2009, \apj, 700, 331,
  \dodoi{10.1088/0004-637X/700/1/331}

\bibitem[{{Hicken} {et~al.}(2012){Hicken}, {Challis}, {Kirshner}, {Rest},
  {Cramer}, {Wood-Vasey}, {Bakos}, {Berlind}, {Brown}, {Caldwell}, {Calkins},
  {Currie}, {de Kleer}, {Esquerdo}, {Everett}, {Falco}, {Fernandez},
  {Friedman}, {Groner}, {Hartman}, {Holman}, {Hutchins}, {Keys}, {Kipping},
  {Latham}, {Marion}, {Narayan}, {Pahre}, {Pal}, {Peters}, {Perumpilly},
  {Ripman}, {Sipocz}, {Szentgyorgyi}, {Tang}, {Torres}, {Vaz}, {Wolk}, \&
  {Zezas}}]{CFA4}
{Hicken}, M., {Challis}, P., {Kirshner}, R.~P., {et~al.} 2012, \apjs, 200, 12,
  \dodoi{10.1088/0067-0049/200/2/12}

\bibitem[{{Hubble}(1929)}]{hubble29}
{Hubble}, E. 1929, Proceedings of the National Academy of Science, 15, 168,
  \dodoi{10.1073/pnas.15.3.168}

\bibitem[{{Huchra}(1992)}]{huchra92}
{Huchra}, J.~P. 1992, Science, 256, 321, \dodoi{10.1126/science.256.5055.321}

\bibitem[{{Jensen} {et~al.}(2015){Jensen}, {Blakeslee}, {Gibson}, {Lee},
  {Cantiello}, {Raimondo}, {Boyer}, \& {Cho}}]{jensen15}
{Jensen}, J.~B., {Blakeslee}, J.~P., {Gibson}, Z., {et~al.} 2015, \apj, 808,
  91, \dodoi{10.1088/0004-637X/808/1/91}

\bibitem[{{Jensen} {et~al.}(2001){Jensen}, {Tonry}, {Thompson}, {Ajhar},
  {Lauer}, {Rieke}, {Postman}, \& {Liu}}]{jensen01}
{Jensen}, J.~B., {Tonry}, J.~L., {Thompson}, R.~I., {et~al.} 2001, \apj, 550,
  503, \dodoi{10.1086/319819}

\bibitem[{{Jensen} {et~al.}(2021){Jensen}, {Blakeslee}, {Ma}, {Milne}, {Brown},
  {Cantiello}, {Garnavich}, {Greene}, {Lucey}, {Phan}, {Tully}, \&
  {Wood}}]{jensen21}
{Jensen}, J.~B., {Blakeslee}, J.~P., {Ma}, C.-P., {et~al.} 2021, \apjs, 255,
  21, \dodoi{10.3847/1538-4365/ac01e7}

\bibitem[{{Jha} {et~al.}(2007){Jha}, {Riess}, \& {Kirshner}}]{jha07}
{Jha}, S., {Riess}, A.~G., \& {Kirshner}, R.~P. 2007, \apj, 659, 122,
  \dodoi{10.1086/512054}

\bibitem[{{Jha} {et~al.}(2006){Jha}, {Kirshner}, {Challis}, {Garnavich},
  {Matheson}, {Soderberg}, {Graves}, {Hicken}, {Alves}, {Arce}, {Balog},
  {Barmby}, {Barton}, {Berlind}, {Bragg}, {Brice{\~n}o}, {Brown}, {Buckley},
  {Caldwell}, {Calkins}, {Carter}, {Concannon}, {Donnelly}, {Eriksen},
  {Fabricant}, {Falco}, {Fiore}, {Garcia}, {G{\'o}mez}, {Grogin}, {Groner},
  {Groot}, {Haisch}, {Hartmann}, {Hergenrother}, {Holman}, {Huchra},
  {Jayawardhana}, {Jerius}, {Kannappan}, {Kim}, {Kleyna}, {Kochanek},
  {Koranyi}, {Krockenberger}, {Lada}, {Luhman}, {Luu}, {Macri}, {Mader},
  {Mahdavi}, {Marengo}, {Marsden}, {McLeod}, {McNamara}, {Megeath}, {Moraru},
  {Mossman}, {Muench}, {Mu{\~n}oz}, {Muzerolle}, {Naranjo}, {Nelson-Patel},
  {Pahre}, {Patten}, {Peters}, {Peters}, {Raymond}, {Rines}, {Schild},
  {Sobczak}, {Spahr}, {Stauffer}, {Stefanik}, {Szentgyorgyi}, {Tollestrup},
  {V{\"a}is{\"a}nen}, {Vikhlinin}, {Wang}, {Willner}, {Wolk}, {Zajac}, {Zhao},
  \& {Stanek}}]{CFA2}
{Jha}, S., {Kirshner}, R.~P., {Challis}, P., {et~al.} 2006, \aj, 131, 527,
  \dodoi{10.1086/497989}

\bibitem[{{Kawabata} {et~al.}(2020){Kawabata}, {Maeda}, {Yamanaka}, {Nakaoka},
  {Kawabata}, {Adachi}, {Akitaya}, {Burgaz}, {Hanayama}, {Horiuchi},
  {Hosokawa}, {Iida}, {Imazato}, {Isogai}, {Jiang}, {Katoh}, {Kimura}, {Kino},
  {Kuroda}, {Maehara}, {Matsubayashi}, {Morihana}, {Murata}, {Nagao}, {Niwano},
  {Nogami}, {Oeda}, {Ono}, {Onozato}, {Otsuka}, {Saito}, {Sasada}, {Shiraishi},
  {Sugiyama}, {Taguchi}, {Takahashi}, {Takagi}, {Takagi}, {Takayama}, {Tozuka},
  \& {Sekiguchi}}]{LOWZ}
{Kawabata}, M., {Maeda}, K., {Yamanaka}, M., {et~al.} 2020, \apj, 893, 143,
  \dodoi{10.3847/1538-4357/ab8236}

\bibitem[{{Kelly} {et~al.}(2010){Kelly}, {Hicken}, {Burke}, {Mandel}, \&
  {Kirshner}}]{kelly10}
{Kelly}, P.~L., {Hicken}, M., {Burke}, D.~L., {Mandel}, K.~S., \& {Kirshner},
  R.~P. 2010, \apj, 715, 743, \dodoi{10.1088/0004-637X/715/2/743}

\bibitem[{{Kelsey} {et~al.}(2021){Kelsey}, {Sullivan}, {Smith}, {Wiseman},
  {Brout}, {Davis}, {Frohmaier}, {Galbany}, {Grayling}, {Guti{\'e}rrez},
  {Hinton}, {Kessler}, {Lidman}, {M{\"o}ller}, {Sako}, {Scolnic}, {Uddin},
  {Vincenzi}, {Abbott}, {Aguena}, {Allam}, {Annis}, {Avila}, {Bacon}, {Bertin},
  {Brooks}, {Burke}, {Carnero Rosell}, {Carrasco Kind}, {Carretero},
  {Castander}, {Costanzi}, {da Costa}, {Desai}, {Diehl}, {Doel}, {Everett},
  {Ferrero}, {Fert{\'e}}, {Flaugher}, {Fosalba}, {Garc{\'\i}a-Bellido},
  {Gerdes}, {Gruen}, {Gruendl}, {Gschwend}, {Gutierrez}, {Hollowood},
  {Honscheid}, {James}, {Kim}, {Kuehn}, {Kuropatkin}, {Lahav}, {Lima},
  {Marshall}, {Martini}, {Menanteau}, {Miquel}, {Morgan}, {Ogando}, {Palmese},
  {Paz-Chinch{\'o}n}, {Plazas}, {Romer}, {S{\'a}nchez}, {Sanchez}, {Serrano},
  {Sevilla-Noarbe}, {Suchyta}, {Tarle}, {Thomas}, {To}, {Varga}, {Walker},
  {Wilkinson}, \& {DES Collaboration}}]{kelsey21}
{Kelsey}, L., {Sullivan}, M., {Smith}, M., {et~al.} 2021, \mnras, 501, 4861,
  \dodoi{10.1093/mnras/staa3924}

\bibitem[{{Kessler} {et~al.}(2009){Kessler}, {Bernstein}, {Cinabro}, {Dilday},
  {Frieman}, {Jha}, {Kuhlmann}, {Miknaitis}, {Sako}, {Taylor}, \&
  {Vanderplas}}]{kessler09}
{Kessler}, R., {Bernstein}, J.~P., {Cinabro}, D., {et~al.} 2009, \pasp, 121,
  1028, \dodoi{10.1086/605984}

\bibitem[{{Khetan} {et~al.}(2021){Khetan}, {Izzo}, {Branchesi}, {Wojtak},
  {Cantiello}, {Murugeshan}, {Agnello}, {Cappellaro}, {Della Valle}, {Gall},
  {Hjorth}, {Benetti}, {Brocato}, {Burke}, {Hiramatsu}, {Howell}, {Tomasella},
  \& {Valenti}}]{khetan21}
{Khetan}, N., {Izzo}, L., {Branchesi}, M., {et~al.} 2021, \aap, 647, A72,
  \dodoi{10.1051/0004-6361/202039196}

\bibitem[{{Lampeitl} {et~al.}(2010){Lampeitl}, {Smith}, {Nichol}, {Bassett},
  {Cinabro}, {Dilday}, {Foley}, {Frieman}, {Garnavich}, {Goobar}, {Im}, {Jha},
  {Marriner}, {Miquel}, {Nordin}, {{\"O}stman}, {Riess}, {Sako}, {Schneider},
  {Sollerman}, \& {Stritzinger}}]{lampeitl10}
{Lampeitl}, H., {Smith}, M., {Nichol}, R.~C., {et~al.} 2010, \apj, 722, 566,
  \dodoi{10.1088/0004-637X/722/1/566}

\bibitem[{{Lema{\^\i}tre}(1927)}]{lemaitre27}
{Lema{\^\i}tre}, G. 1927, Annales de la Soci\&eacute;t\&eacute; Scientifique de
  Bruxelles, 47, 49

\bibitem[{{Ma} {et~al.}(2014){Ma}, {Greene}, {McConnell}, {Janish},
  {Blakeslee}, {Thomas}, \& {Murphy}}]{ma14}
{Ma}, C.-P., {Greene}, J.~E., {McConnell}, N., {et~al.} 2014, \apj, 795, 158,
  \dodoi{10.1088/0004-637X/795/2/158}

\bibitem[{{Mosher} {et~al.}(2014){Mosher}, {Guy}, {Kessler}, {Astier},
  {Marriner}, {Betoule}, {Sako}, {El-Hage}, {Biswas}, {Pain}, {Kuhlmann},
  {Regnault}, {Frieman}, \& {Schneider}}]{mosher14}
{Mosher}, J., {Guy}, J., {Kessler}, R., {et~al.} 2014, \apj, 793, 16,
  \dodoi{10.1088/0004-637X/793/1/16}

\bibitem[{{Perlmutter} {et~al.}(1998){Perlmutter}, {Aldering}, {della Valle},
  {Deustua}, {Ellis}, {Fabbro}, {Fruchter}, {Goldhaber}, {Groom}, {Hook},
  {Kim}, {Kim}, {Knop}, {Lidman}, {McMahon}, {Nugent}, {Pain}, {Panagia},
  {Pennypacker}, {Ruiz-Lapuente}, {Schaefer}, \& {Walton}}]{perlmutter98}
{Perlmutter}, S., {Aldering}, G., {della Valle}, M., {et~al.} 1998, \nat, 391,
  51, \dodoi{10.1038/34124}

\bibitem[{{Perlmutter} {et~al.}(1999){Perlmutter}, {Aldering}, {Goldhaber},
  {Knop}, {Nugent}, {Castro}, {Deustua}, {Fabbro}, {Goobar}, {Groom}, {Hook},
  {Kim}, {Kim}, {Lee}, {Nunes}, {Pain}, {Pennypacker}, {Quimby}, {Lidman},
  {Ellis}, {Irwin}, {McMahon}, {Ruiz-Lapuente}, {Walton}, {Schaefer}, {Boyle},
  {Filippenko}, {Matheson}, {Fruchter}, {Panagia}, {Newberg}, {Couch}, \&
  {Project}}]{perlmutter99}
{Perlmutter}, S., {Aldering}, G., {Goldhaber}, G., {et~al.} 1999, \apj, 517,
  565, \dodoi{10.1086/307221}

\bibitem[{{Phillips}(1993)}]{phillips93}
{Phillips}, M.~M. 1993, \apjl, 413, L105, \dodoi{10.1086/186970}

\bibitem[{{Phillips} {et~al.}(1999){Phillips}, {Lira}, {Suntzeff}, {Schommer},
  {Hamuy}, \& {Maza}}]{phillips99}
{Phillips}, M.~M., {Lira}, P., {Suntzeff}, N.~B., {et~al.} 1999, \aj, 118,
  1766, \dodoi{10.1086/301032}

\bibitem[{{Pietrzy{\'n}ski} {et~al.}(2019){Pietrzy{\'n}ski}, {Graczyk},
  {Gallenne}, {Gieren}, {Thompson}, {Pilecki}, {Karczmarek}, {G{\'o}rski},
  {Suchomska}, {Taormina}, {Zgirski}, {Wielg{\'o}rski}, {Ko{\l}aczkowski},
  {Konorski}, {Villanova}, {Nardetto}, {Kervella}, {Bresolin}, {Kudritzki},
  {Storm}, {Smolec}, \& {Narloch}}]{LMC2019}
{Pietrzy{\'n}ski}, G., {Graczyk}, D., {Gallenne}, A., {et~al.} 2019, \nat, 567,
  200, \dodoi{10.1038/s41586-019-0999-4}

\bibitem[{{Planck Collaboration} {et~al.}(2020){Planck Collaboration},
  {Aghanim}, {Akrami}, {Ashdown}, {Aumont}, {Baccigalupi}, {Ballardini},
  {Banday}, {Barreiro}, {Bartolo}, {Basak}, {Battye}, {Benabed}, {Bernard},
  {Bersanelli}, {Bielewicz}, {Bock}, {Bond}, {Borrill}, {Bouchet}, {Boulanger},
  {Bucher}, {Burigana}, {Butler}, {Calabrese}, {Cardoso}, {Carron},
  {Challinor}, {Chiang}, {Chluba}, {Colombo}, {Combet}, {Contreras}, {Crill},
  {Cuttaia}, {de Bernardis}, {de Zotti}, {Delabrouille}, {Delouis}, {Di
  Valentino}, {Diego}, {Dor{\'e}}, {Douspis}, {Ducout}, {Dupac}, {Dusini},
  {Efstathiou}, {Elsner}, {En{\ss}lin}, {Eriksen}, {Fantaye}, {Farhang},
  {Fergusson}, {Fernandez-Cobos}, {Finelli}, {Forastieri}, {Frailis},
  {Fraisse}, {Franceschi}, {Frolov}, {Galeotta}, {Galli}, {Ganga},
  {G{\'e}nova-Santos}, {Gerbino}, {Ghosh}, {Gonz{\'a}lez-Nuevo}, {G{\'o}rski},
  {Gratton}, {Gruppuso}, {Gudmundsson}, {Hamann}, {Handley}, {Hansen},
  {Herranz}, {Hildebrandt}, {Hivon}, {Huang}, {Jaffe}, {Jones}, {Karakci},
  {Keih{\"a}nen}, {Keskitalo}, {Kiiveri}, {Kim}, {Kisner}, {Knox},
  {Krachmalnicoff}, {Kunz}, {Kurki-Suonio}, {Lagache}, {Lamarre}, {Lasenby},
  {Lattanzi}, {Lawrence}, {Le Jeune}, {Lemos}, {Lesgourgues}, {Levrier},
  {Lewis}, {Liguori}, {Lilje}, {Lilley}, {Lindholm}, {L{\'o}pez-Caniego},
  {Lubin}, {Ma}, {Mac{\'\i}as-P{\'e}rez}, {Maggio}, {Maino}, {Mandolesi},
  {Mangilli}, {Marcos-Caballero}, {Maris}, {Martin}, {Martinelli},
  {Mart{\'\i}nez-Gonz{\'a}lez}, {Matarrese}, {Mauri}, {McEwen}, {Meinhold},
  {Melchiorri}, {Mennella}, {Migliaccio}, {Millea}, {Mitra},
  {Miville-Desch{\^e}nes}, {Molinari}, {Montier}, {Morgante}, {Moss}, {Natoli},
  {N{\o}rgaard-Nielsen}, {Pagano}, {Paoletti}, {Partridge}, {Patanchon},
  {Peiris}, {Perrotta}, {Pettorino}, {Piacentini}, {Polastri}, {Polenta},
  {Puget}, {Rachen}, {Reinecke}, {Remazeilles}, {Renzi}, {Rocha}, {Rosset},
  {Roudier}, {Rubi{\~n}o-Mart{\'\i}n}, {Ruiz-Granados}, {Salvati}, {Sandri},
  {Savelainen}, {Scott}, {Shellard}, {Sirignano}, {Sirri}, {Spencer},
  {Sunyaev}, {Suur-Uski}, {Tauber}, {Tavagnacco}, {Tenti}, {Toffolatti},
  {Tomasi}, {Trombetti}, {Valenziano}, {Valiviita}, {Van Tent}, {Vibert},
  {Vielva}, {Villa}, {Vittorio}, {Wandelt}, {Wehus}, {White}, {White},
  {Zacchei}, \& {Zonca}}]{planck20}
{Planck Collaboration}, {Aghanim}, N., {Akrami}, Y., {et~al.} 2020, \aap, 641,
  A6, \dodoi{10.1051/0004-6361/201833910}

\bibitem[{{Pruzhinskaya} {et~al.}(2020){Pruzhinskaya}, {Novinskaya}, {Pauna},
  \& {Rosnet}}]{pruzhinskaya20}
{Pruzhinskaya}, M.~V., {Novinskaya}, A.~K., {Pauna}, N., \& {Rosnet}, P. 2020,
  \mnras, 499, 5121, \dodoi{10.1093/mnras/staa3173}

\bibitem[{{Riess} {et~al.}(2019){Riess}, {Casertano}, {Yuan}, {Macri}, \&
  {Scolnic}}]{riess19}
{Riess}, A.~G., {Casertano}, S., {Yuan}, W., {Macri}, L.~M., \& {Scolnic}, D.
  2019, \apj, 876, 85, \dodoi{10.3847/1538-4357/ab1422}

\bibitem[{{Riess} {et~al.}(1996){Riess}, {Press}, \& {Kirshner}}]{rpk96}
{Riess}, A.~G., {Press}, W.~H., \& {Kirshner}, R.~P. 1996, \apj, 473, 88,
  \dodoi{10.1086/178129}

\bibitem[{{Riess} {et~al.}(1998){Riess}, {Filippenko}, {Challis},
  {Clocchiatti}, {Diercks}, {Garnavich}, {Gilliland}, {Hogan}, {Jha},
  {Kirshner}, {Leibundgut}, {Phillips}, {Reiss}, {Schmidt}, {Schommer},
  {Smith}, {Spyromilio}, {Stubbs}, {Suntzeff}, \& {Tonry}}]{riess98}
{Riess}, A.~G., {Filippenko}, A.~V., {Challis}, P., {et~al.} 1998, \aj, 116,
  1009, \dodoi{10.1086/300499}

\bibitem[{{Riess} {et~al.}(1999){Riess}, {Kirshner}, {Schmidt}, {Jha},
  {Challis}, {Garnavich}, {Esin}, {Carpenter}, {Grashius}, {Schild}, {Berlind},
  {Huchra}, {Prosser}, {Falco}, {Benson}, {Brice{\~n}o}, {Brown}, {Caldwell},
  {dell'Antonio}, {Filippenko}, {Goodman}, {Grogin}, {Groner}, {Hughes},
  {Green}, {Jansen}, {Kleyna}, {Luu}, {Macri}, {McLeod}, {McLeod}, {McNamara},
  {McLean}, {Milone}, {Mohr}, {Moraru}, {Peng}, {Peters}, {Prestwich},
  {Stanek}, {Szentgyorgyi}, \& {Zhao}}]{CFA1}
{Riess}, A.~G., {Kirshner}, R.~P., {Schmidt}, B.~P., {et~al.} 1999, \aj, 117,
  707, \dodoi{10.1086/300738}

\bibitem[{{Riess} {et~al.}(2009){Riess}, {Macri}, {Casertano}, {Sosey},
  {Lampeitl}, {Ferguson}, {Filippenko}, {Jha}, {Li}, {Chornock}, \&
  {Sarkar}}]{riess09}
{Riess}, A.~G., {Macri}, L., {Casertano}, S., {et~al.} 2009, \apj, 699, 539,
  \dodoi{10.1088/0004-637X/699/1/539}

\bibitem[{{Riess} {et~al.}(2016){Riess}, {Macri}, {Hoffmann}, {Scolnic},
  {Casertano}, {Filippenko}, {Tucker}, {Reid}, {Jones}, {Silverman},
  {Chornock}, {Challis}, {Yuan}, {Brown}, \& {Foley}}]{riess16}
{Riess}, A.~G., {Macri}, L.~M., {Hoffmann}, S.~L., {et~al.} 2016, \apj, 826,
  56, \dodoi{10.3847/0004-637X/826/1/56}

\bibitem[{{Riess} {et~al.}(2022){Riess}, {Yuan}, {Macri}, {Scolnic}, {Brout},
  {Casertano}, {Jones}, {Murakami}, {Anand}, {Breuval}, {Brink}, {Filippenko},
  {Hoffmann}, {Jha}, {D'arcy Kenworthy}, {Mackenty}, {Stahl}, \&
  {Zheng}}]{riess21}
{Riess}, A.~G., {Yuan}, W., {Macri}, L.~M., {et~al.} 2022, \apjl, 934, L7,
  \dodoi{10.3847/2041-8213/ac5c5b}

\bibitem[{{Rose} {et~al.}(2019){Rose}, {Garnavich}, \& {Berg}}]{rose19}
{Rose}, B.~M., {Garnavich}, P.~M., \& {Berg}, M.~A. 2019, \apj, 874, 32,
  \dodoi{10.3847/1538-4357/ab0704}

\bibitem[{{Scannapieco} \& {Bildsten}(2005)}]{scannapieco05}
{Scannapieco}, E., \& {Bildsten}, L. 2005, \apjl, 629, L85,
  \dodoi{10.1086/452632}

\bibitem[{{Scolnic} {et~al.}(2020){Scolnic}, {Smith}, {Massiah}, {Wiseman},
  {Brout}, {Kessler}, {Davis}, {Foley}, {Galbany}, {Hinton}, {Hounsell},
  {Kelsey}, {Lidman}, {Macaulay}, {Morgan}, {Nichol}, {M{\"o}ller}, {Popovic},
  {Sako}, {Sullivan}, {Thomas}, {Tucker}, {Abbott}, {Aguena}, {Allam}, {Annis},
  {Avila}, {Bechtol}, {Bertin}, {Brooks}, {Burke}, {Rosell}, {Carollo}, {Kind},
  {Carretero}, {Costanzi}, {da Costa}, {De Vicente}, {Desai}, {Diehl}, {Doel},
  {Drlica-Wagner}, {Eckert}, {Eifler}, {Everett}, {Flaugher}, {Fosalba},
  {Frieman}, {Garc{\'\i}a-Bellido}, {Gaztanaga}, {Gerdes}, {Glazebrook},
  {Gruen}, {Gruendl}, {Gschwend}, {Gutierrez}, {Hartley}, {Hollowood},
  {Honscheid}, {James}, {Kuehn}, {Kuropatkin}, {Lewis}, {Li}, {Lima}, {Maia},
  {Marshall}, {Menanteau}, {Miquel}, {Palmese}, {Paz-Chinch{\'o}n}, {Plazas},
  {Pursiainen}, {Sanchez}, {Scarpine}, {Schubnell}, {Serrano},
  {Sevilla-Noarbe}, {Sommer}, {Suchyta}, {Swanson}, {Tarle}, {Varga}, {Walker},
  {Wilkinson}, \& {DES Collaboration}}]{scolnic20}
{Scolnic}, D., {Smith}, M., {Massiah}, A., {et~al.} 2020, \apjl, 896, L13,
  \dodoi{10.3847/2041-8213/ab8735}

\bibitem[{{Scolnic} {et~al.}(2022){Scolnic}, {Brout}, {Carr}, {Riess}, {Davis},
  {Dwomoh}, {Jones}, {Ali}, {Charvu}, {Chen}, {Peterson}, {Popovic}, {Rose},
  {Wood}, {Brown}, {Chambers}, {Coulter}, {Dettman}, {Dimitriadis},
  {Filippenko}, {Foley}, {Jha}, {Kilpatrick}, {Kirshner}, {Pan}, {Rest},
  {Rojas-Bravo}, {Siebert}, {Stahl}, \& {Zheng}}]{pantheon+1}
{Scolnic}, D., {Brout}, D., {Carr}, A., {et~al.} 2022, \apj, 938, 113,
  \dodoi{10.3847/1538-4357/ac8b7a}

\bibitem[{{Sullivan} {et~al.}(2006){Sullivan}, {Le Borgne}, {Pritchet},
  {Hodsman}, {Neill}, {Howell}, {Carlberg}, {Astier}, {Aubourg}, {Balam},
  {Basa}, {Conley}, {Fabbro}, {Fouchez}, {Guy}, {Hook}, {Pain},
  {Palanque-Delabrouille}, {Perrett}, {Regnault}, {Rich}, {Taillet}, {Baumont},
  {Bronder}, {Ellis}, {Filiol}, {Lusset}, {Perlmutter}, {Ripoche}, \&
  {Tao}}]{sullivan06}
{Sullivan}, M., {Le Borgne}, D., {Pritchet}, C.~J., {et~al.} 2006, \apj, 648,
  868, \dodoi{10.1086/506137}

\bibitem[{{Sullivan} {et~al.}(2010){Sullivan}, {Conley}, {Howell}, {Neill},
  {Astier}, {Balland}, {Basa}, {Carlberg}, {Fouchez}, {Guy}, {Hardin}, {Hook},
  {Pain}, {Palanque-Delabrouille}, {Perrett}, {Pritchet}, {Regnault}, {Rich},
  {Ruhlmann-Kleider}, {Baumont}, {Hsiao}, {Kronborg}, {Lidman}, {Perlmutter},
  \& {Walker}}]{sullivan10}
{Sullivan}, M., {Conley}, A., {Howell}, D.~A., {et~al.} 2010, \mnras, 406, 782,
  \dodoi{10.1111/j.1365-2966.2010.16731.x}

\bibitem[{{Taylor} {et~al.}(2021){Taylor}, {Lidman}, {Tucker}, {Brout},
  {Hinton}, \& {Kessler}}]{taylor21}
{Taylor}, G., {Lidman}, C., {Tucker}, B.~E., {et~al.} 2021, \mnras, 504, 4111,
  \dodoi{10.1093/mnras/stab962}

\bibitem[{{Tonry} \& {Schneider}(1988)}]{tonry88}
{Tonry}, J., \& {Schneider}, D.~P. 1988, \aj, 96, 807, \dodoi{10.1086/114847}

\bibitem[{{Tripp}(1998)}]{tripp98}
{Tripp}, R. 1998, \aap, 331, 815

\end{thebibliography}

\end{document}